\documentclass[%
 reprint,
superscriptaddress,
 amsmath,amssymb,
 aps,twocolumn,
]{revtex4-2}

\usepackage{graphicx}
\usepackage{dcolumn}
\usepackage{bm}
\usepackage[hypertexnames=false]{hyperref}
\usepackage{siunitx}
\usepackage[capitalise]{cleveref}
\usepackage{braket}

\newcommand{\sigmah}{\hat{\sigma}}
\newcommand{\ah}{\hat{a}}

\usepackage{physics} 
\usepackage{enumerate}
\usepackage{dsfont}

\newcommand{\ds}{\chi''(\Delta_{\mathrm{pa}})}
\newcommand{\dsa}{\chi_\mathrm{a}(\Delta_{\mathrm{pa}})}

\newcommand{\renyi}{R\'{e}nyi }

\usepackage{xcolor}

\usepackage[normalem]{ulem}

\begin{document}

\title{Engineering random spin models with atoms in a high-finesse cavity}

\author{Nick Sauerwein}
\author{Francesca Orsi}
\affiliation{Institute of Physics and Center for Quantum Science and Engineering, \'Ecole Polytechnique F\'ed\'erale de Lausanne (EPFL), CH-1015 Lausanne, Switzerland}%
\author{Philipp Uhrich}
\author{Soumik Bandyopadhyay}
\affiliation{INO-CNR BEC Center \& Department of Physics, University of Trento, Via Sommarive 14, I-38123 Trento, Italy}%
\affiliation{INFN-TIFPA, Trento Institute for Fundamental Physics and Applications, Trento, Italy}

\author{Francesco Mattiotti}
\affiliation{University of Strasbourg and CNRS, CESQ and ISIS (UMR 7006), aQCess, 67000 Strasbourg, France}%

\author{Tigrane Cantat-Moltrecht}
\affiliation{Institute of Physics and Center for Quantum Science and Engineering, \'Ecole Polytechnique F\'ed\'erale de Lausanne (EPFL), CH-1015 Lausanne, Switzerland}%

\author{Guido Pupillo}
\affiliation{University of Strasbourg and CNRS, CESQ and ISIS (UMR 7006), aQCess, 67000 Strasbourg, France}%

\author{Philipp Hauke}
\affiliation{INO-CNR BEC Center \& Department of Physics, University of Trento, Via Sommarive 14, I-38123 Trento, Italy}%
\affiliation{INFN-TIFPA, Trento Institute for Fundamental Physics and Applications, Trento, Italy}
\author{Jean-Philippe Brantut}%
\email{jean-philippe.brantut@epfl.ch}
\affiliation{Institute of Physics and Center for Quantum Science and Engineering, \'Ecole Polytechnique F\'ed\'erale de Lausanne (EPFL), CH-1015 Lausanne, Switzerland}%
\date{\today}

\begin{abstract}
    All-to-all interacting, disordered quantum many-body models have a wide range of applications across disciplines, from spin glasses in condensed-matter physics, over holographic duality in high-energy physics, to annealing algorithms in quantum computing. Typically, these models are abstractions that do not find unambiguous physical realisations in nature. 
    Here, we realise an all-to-all interacting, disordered spin system by subjecting an atomic cloud in a cavity to a controllable light shift. 
    Adjusting the detuning between atom resonance and cavity mode, we can tune between disordered versions of a central-mode model and a Lipkin--Meshkov--Glick model. 
    By spectroscopically probing the low-energy excitations of the system, we explore the competition of interactions with disorder across a broad parameter range. 
    We show how disorder in the central-mode model breaks the strong collective coupling, making the dark state manifold cross over to a random distribution of weakly-mixed light--matter, “grey”, states. 
    In the Lipkin--Meshkov--Glick model the ferromagnetic finite-size ground state evolves towards a paramagnet as disorder is increased. 
    In that regime, semi-localised eigenstates emerge, as we observe by extracting bounds on the participation ratio. 
    These results present significant steps towards freely programmable cavity-mediated interactions for the design of arbitrary spin Hamiltonians.    
\end{abstract}

\maketitle

\begin{figure}
	\includegraphics{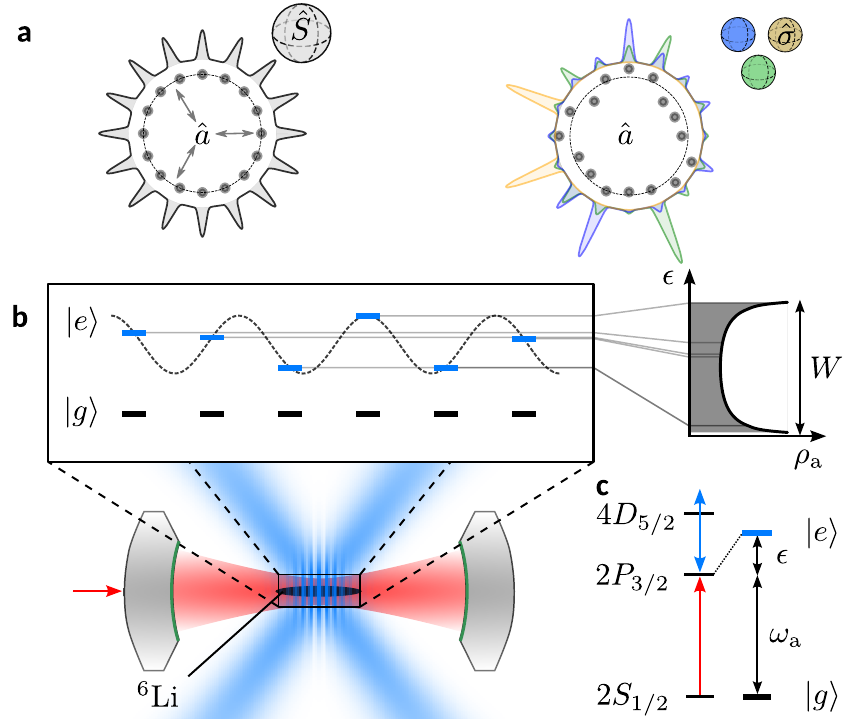}
	\caption{\label{fig:intro} \textbf{Concept of the experiment.} \textbf{a}, Fragmentation of collective light--matter eigenstates with increasing disorder.
    Left: Disorder-free system with all spins (spheres
    ) identically coupled to the central mode $\hat{a}$ provided by the cavity field, forming a symmetric collective Dicke state. 
    Right: With disorder, the collective state fragments into few- or single-spin ensembles whose constituents are located at arbitrarily large distances, exchanging excitations through the cavity, sketched here for three excitation modes.
    \textbf{b} Experimental realisation: Atoms are trapped in an optical resonator, forming an atom array commensurate with the cavity mode, ensuring identical atom--light coupling. 
    Two crossed light-shifting beams (blue) illuminate the atoms with an incommensurate standing-wave inference pattern, leading to a quasi-random intensity distribution $\rho_\mathrm{a}$ over the atoms (right).
    \textbf{c}, Simplified level-diagram of the $^6$Li atoms.
    The light-shifting laser (blue arrow) off-resonantly couples the $2P_{3/2}$ manifold with the higher-lying $4D_{5/2}$ manifold, yielding a dressed state $\ket{e}$ (blue), with  an energy shift proportional to the laser intensity.}
\end{figure}

The unavoidable presence of impurities and inhomogeneities in most real-world physical systems has given a strong motivation to the study of disordered models.  
In such studies, important insights into the typical behaviour of a many-body system can be obtained by considering an ensemble of realisations with randomly distributed parameters \cite{parisi}. 
In this way, a deeper understanding of the structure of low-energy excitations in complex quantum systems can be achieved, providing keys to interpreting transport and thermodynamics observations. 
Going one step further, several quantum simulation platforms, such as trapped ions \cite{blattQuantumSimulationsTrapped2012}, ultracold atoms \cite{grossQuantumSimulationsUltracold2017} and Rydberg atoms \cite{lippe:2021wi,signoles:2021us,marcuzzi:2017vh}, have demonstrated the capability to implement controlled disorder into otherwise clean many-body systems. Those allowed for the investigation of non-equilibrium dynamics, revealing some of the most intriguing phenomena of random systems, such as Anderson \cite{roatiAndersonLocalizationNoninteracting2008, billyDirectObservationAnderson2008,jendrzejewskiThreedimensionalLocalizationUltracold2012,kondovThreeDimensionalAndersonLocalization2011, maierEnvironmentAssistedQuantumTransport2019} and many-body localisation \cite{schreiberObservationManybodyLocalization2015,smithManybodyLocalizationQuantum2016,lukinProbingEntanglementManybody2019}.

In the last years, cavity quantum electrodynamics (QED) has emerged as a new platform for quantum simulation.
By harnessing photons to tailor novel types of interactions beyond the native van~der~Waals and dipolar interactions between atoms, 
cavity QED unites the scalability of atom devices with tunable long-range interactions \cite{mivehvarCavityQEDQuantum2021}. 
Previous experiments used this platform to explore new, superradiant \cite{Baumann:2010aa,guo:2021aa,zhang:2021tr,Leonard:2017aa} as well as dissipation-stabilised \cite{Dogra:2019aa,kongkhambut:wk} phases of matter in quantum gases, and to demonstrate tunable-range interactions \cite{Vaidya:2018aa} and emergent geometries using spatial and spectral addressing \cite{periwalProgrammableInteractionsEmergent2021}.

In this article, we implement random spin models on a cavity QED platform and study their low-lying excitations.
Via a light-shift technique, we realise a quasi-random longitudinal field with controlled strength, which competes with an all-to-all flip-flop interaction mediated by the exchange of cavity photons. 
Leveraging the open nature of the cavity, we observe the frequency-resolved response in the cavity field and atomic polarisation channels. 
We exploit our setup to observe disorder-driven crossovers in two different regimes: a central-mode model where we observe a disorder-induced dressing of otherwise dark anti-symmetric states with cavity photons, and a Lipkin--Meshkov--Glick model (an instance of a Richardson--Gaudin model) where disorder competes with ferromagnetic order. As we show theoretically and experimentally, the frequency-resolved susceptibilities are sensitive to the detailed structure of excitations, providing insights in particular about their localisation properties. Our system is a natural starting point to investigate the spectacular consequences of strong light–matter coupling on materials properties  \cite{ebbesenHybridLightMatterStates2016, garcia-vidalManipulatingMatterStrong2021,
blochStronglyCorrelatedElectronPhoton2022} such as transport \cite{orgiuConductivityOrganicSemiconductors2015, lerarioHighSpeedFlowOrganicPolaritons2017, appuglieseBreakdownTopologicalProtection2022} or magnetism \cite{thomasLargeEnhancementFerromagnatismNanoparticles}, where the effect of disorder due to impurities and material inhomogeneities is believed to be strongly influenced by light.

\begin{figure*}
	\includegraphics{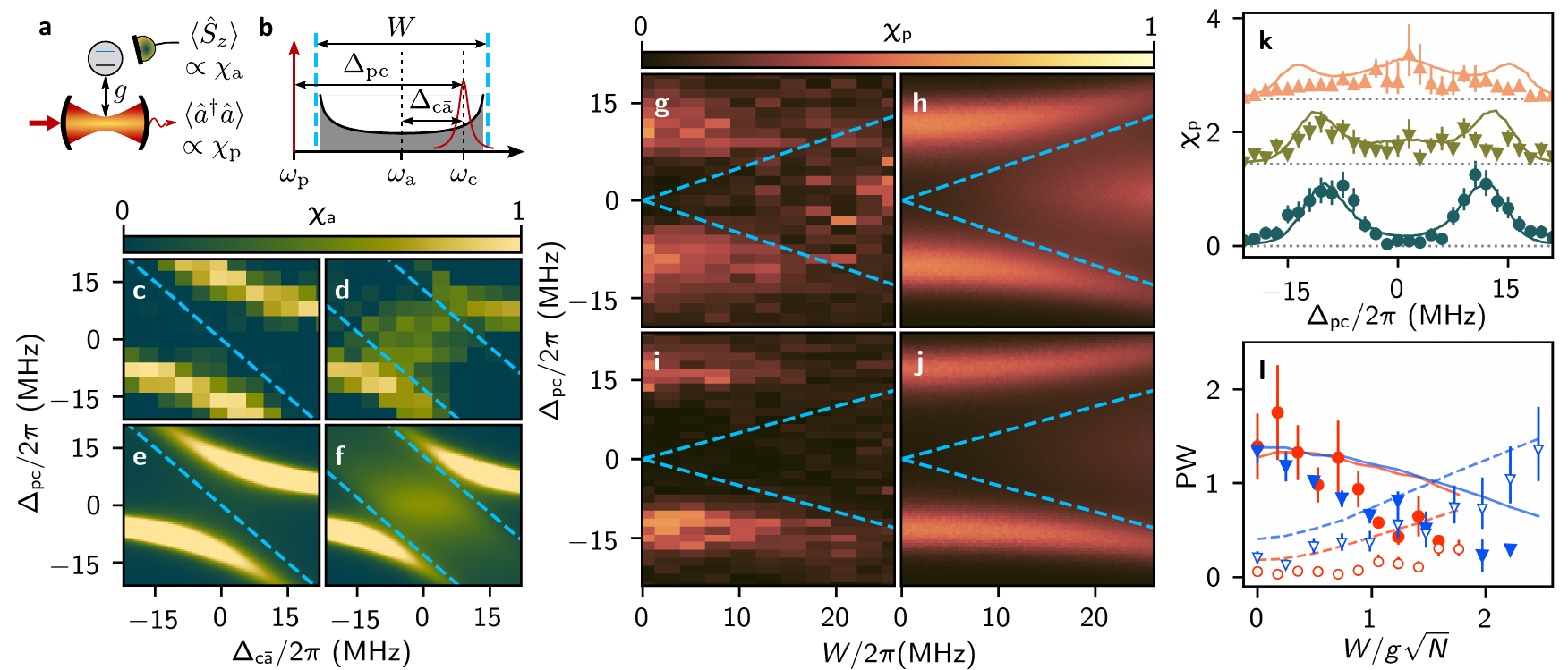}
	\caption{\label{fig:RTC} \textbf{Response of the system in the central-mode regime.} 
        \textbf{a}, Measurement of atomic and photonic susceptibilities upon a drive of the cavity.
        The photonic susceptibility $\chi_{\mathrm{p}}$ is given by a cavity transmission measurement, while the atomic susceptibility $\chi_{\mathrm{a}}$ is proportional to the number of atoms that have been excited by the cavity field (see Methods).
        \textbf{b}, Frequency diagram illustrating the relative detunings between the atoms with average frequency $\omega_{\rm \overline{a}}$ in a range $W$, the cavity at frequency $\omega_{\rm c}$, and the probe at $\omega_{\rm p}$.
        The light-blue dashed lines indicate the edges of the atomic frequency distribution.
        In all other panels, the atomic states lie between the two light-blue lines.
        \textbf{c--f}, Measured (\textbf{c}, \textbf{d}) and simulated (\textbf{e}, \textbf{f}, see Methods \cref{s:near-res-theory}, with $N =100 \pm 30$) atomic susceptibility maps as a function of atom--cavity and pump--cavity detunings ($x$- and $y$-axis, respectively), for the clean system (\textbf{c}, \textbf{e}) and at maximal disorder $W/2\pi = \SI{26}{\MHz}$ (\textbf{d}, \textbf{f}). \textbf{g--j}, Measured (\textbf{g}, \textbf{i}) and simulated (\textbf{h}, \textbf{j}) photonic susceptibility as a function of disorder strength $W$ for $74\pm 31$ (\textbf{g}, \textbf{h}) and $145\pm43$ (\textbf{i}, \textbf{j}) atoms.   \textbf{k}, Vertical sections of panels \textbf{g}~and~\textbf{h} overlapped with simulations (curves are offset vertically for clarity) for $W/g\sqrt{N}=0,1,2$.
        \textbf{l}, Photon weight PW of the grey states (empty markers, dashed lines) and the polaritons (filled markers, continuous lines) as a function of normalised disorder strength for $ N =145$ (circles) and $ N =74$ (triangles) atoms, indicating the disappearance of the polaritons and the appearance of the grey states.
        The grey states' photon weight was measured by taking the average photonic susceptibility over the grey state region defined by $\Delta_{\mathrm{pc}} \in \{-\Gamma/2, \Gamma/2\}$, while the photon weight of the polariton was quantified by taking the height of the lower polariton in \textbf{g}~and~\textbf{i}, which is not affected by the radiation pressure of the light-shifting beam (see Methods).}
\end{figure*}

\section{Model}
\label{s:model}
Our system implements a paradigmatic model consisting of $N$ Ising spins, mapped to internal atomic states, identically coupled to the central, bosonic photon mode of the cavity.
By exposing the $i$th spin to a random energy shift $\epsilon_{i}$, the model is described by the disordered Tavis--Cummings-type Hamiltonian 
\begin{equation}\label{eq:Hamiltonian}
\hat{H}_\mathrm{TC} =  \Delta_{\mathrm{ca}} \ah^{\dagger}\ah+ g \sqrt{N} \left(\hat{S}^+ \ah + \hat{S}^- \ah^{\dagger} \right) + \sum_{i=1}^N \epsilon_{i} \, \frac{\sigmah_i^{z}}{2}.
\end{equation}
Here, $\ah^{\dagger}$ and $\ah$ are the creation and annihilation operators of photons in the cavity, $\sigmah_i^{r}$ are the $r$-Pauli operators acting on the Ising (pseudo-)spin-$1/2$ of the $i^{th}$ atom, $\hat{S}^{+(-)}=\sum_{i=1}^N \sigmah_i^{+(-)} / \sqrt{N}$ are the collective spin-raising (lowering) operators, and $\Delta_{\mathrm{ca}}$ is the detuning between the cavity and the bare atomic resonance. 
We set $\hbar=1$ throughout the manuscript.
Central mode models \cite{prokofevTheorySpinBath2000, dukelskyExactlySolvableRichardsonGaudin2004} have been used to describe a large variety of physical situations, including qubit decoherence in solid-state quantum computing platforms as well as heat and charge transport in nanostructures. 

In the disorder-free instance of the Hamiltonian of \cref{eq:Hamiltonian} (\cref{fig:intro}\textbf{a}, left), the spin-$1/2$ degrees-of-freedom form a manifold of $N+1$ collective exchange-symmetric Dicke states coupled to light, thus called `bright states', which are described by a single collective spin $\hat{S}$. The remaining $2^N-(N+1)$ states form a dark manifold, which is decoupled from the cavity field. In the single excitation manifold, this structure reduces to two polaritons and $N-1$ dark states. A controlled breaking of this collective spin description into macroscopic subsets that are spatially and spectrally distinguishable has recently been demonstrated by splitting atomic ensembles with the help of optical tweezers and magnetic field gradients~\cite{periwalProgrammableInteractionsEmergent2021}.

In the model of \cref{eq:Hamiltonian}, the collective spin description is broken by disorder, as illustrated in \cref{fig:intro}\textbf{a}, right.
This leads to a fragmentation of the dark state manifold into an ensemble of `grey eigenstates' that are hybridisations of the delocalised photon field and of a few localised spins with similar energies \cite{orgiuConductivityOrganicSemiconductors2015}. Because the coupling to the cavity extends over the entire system, energy resonances between spins can occur at arbitrarily large distances in the presence of disorder. As a result, grey eigenstates have wave functions that are neither localised nor delocalised, but {\it semi-localised} over multiple, arbitrarily distant spins \cite{botzungDarkStateSemilocalization2020, scholesPolaritonsExcitons2020}. 
It was recently demonstrated theoretically that for any strength of light--matter coupling this results in a {\it multi-fractal} structure of the eigenstates, similar to that found at the critical points of localisation--delocalisation transitions~\cite{dubailLargeRandomArrowhead2022}. 
Even though they have never been directly observed, it is believed that disorder-induced grey states are responsible for the spectacular enhancement of energy and charge transport found in disordered molecular systems coupled to cavities \cite{orgiuConductivityOrganicSemiconductors2015, lerarioHighSpeedFlowOrganicPolaritons2017, schachenmayerCavityEnhancedTransportExcitons2015, feistExtraordinaryExcitonConductance2015, gonzalez-ballesteroUncoupledDarkStates2016, chavezDisorderEnhancedDisorderIndepTransport2021}. 

Experimentally, the Hamiltonian in \cref{eq:Hamiltonian} is realised by an array of $N = 90$ to $800$ thermal $^6\mathrm{Li}$ atoms confined in about $160$ trapping sites, positioned at the anti-nodes of the resonant cavity field.
The spins are encoded in the $2S_{1/2}^{F=1/2}$ ($\ket{g}$) and $2P_{3/2}$ ($\ket{e}$) states of $^6$Li atoms (\cref{fig:intro}\textbf{b},~\textbf{c}).
The cavity resonance is tuned close to the $2S_{1/2}$--$2P_{3/2}$ transition at \SI{671}{\nm}, with the detuning given by $\Delta_{\mathrm{ca}}$. Our cavity is close to concentric, leading to a single-atom cooperativity of $\eta = (4g^2)/(\kappa\Gamma) = 6.4$ with $g/2\pi,\kappa/2\pi,\Gamma/2\pi = \SI{2.05}{\MHz},\, \SI{0.45}{\MHz},\,\SI{5.8}{\MHz}$. Due to the cloud's temperature of $\SI{200}{\mu \K}$, and the reduced dipole moment for linearly polarised light at zero magnetic field, the average cavity coupling that the atoms experience is $\bar{g}/2\pi = \SI{1.23}{\MHz}$ (see Methods).

The disorder is created by two laser beams that intersect at an angle of \SI{25.6}{\degree} at the position of the atoms, with frequency slightly detuned from the $2P_{3/2}$--$4D_{5/2}$ transition at \SI{460}{\nm}, forming a light-shifting lattice that is incommensurate with the trapping lattice. This produces a quasi-random pattern of strong light-shifts of the $2P_{3/2}$ state, with negligible effect on atoms in the ground state, as illustrated in \cref{fig:intro}\textbf{b},~\textbf{c}. 
These light-shifts result in quasi-disordered energy shifts $\epsilon_i$, that translate into the spin language as random local longitudinal fields sampled from the distribution $\rho_\mathrm{a}(\epsilon) = \left[\pi \sqrt{\epsilon ( W - \epsilon )} \right]^{-1}$, where $W$ is proportional to the intensity of the control laser (see Methods). 

We probe the system by weakly driving the cavity on-axis with a probe beam and measuring both the photon transmission proportional to $\langle \ah^{\dagger}\ah\rangle$, and the atomic excitations $\langle \hat{S}^z \rangle = \langle \sum_{i=1}^N \sigmah_i^{z} \rangle  / (2N)$ using an optical pumping technique, as presented in \cref{fig:RTC}\textbf{a}. In the linear response regime, this provides us with the frequency-dependent photonic and atomic (spin) susceptibilities, $\chi_{\mathrm{p}}$ and $\chi_{\mathrm{a}}$ (see Eqs.~\eqref{chip_definition_onres},~\eqref{chia_definition_onres}, and~\eqref{e:atomic_susc} of Methods for definitions).

\section{Near-resonant regime and grey states}
\label{s:near-resonant}

We first investigate the regime at small $\Delta_{\mathrm{c}\bar{\mathrm{a}}}$ where the cavity resonance is close to the mean atomic resonances, $\Delta_{\mathrm{c}\bar{\mathrm{a}}} = \Delta_{\mathrm{ca}} - W/2$ (see \cref{fig:RTC}\textbf{b}). In the absence of disorder, we observe the canonical normal-mode splitting of width $2 g \sqrt{N}/2\pi = \SI{22}{\MHz}$ expected from the Tavis--Cummings model, as shown in \cref{fig:RTC}\textbf{c}.
As a result of this splitting, a Rabi gap forms at $\Delta_{\mathrm{c}\bar{\mathrm{a}}} = 0$, and direct atomic excitations at the bare resonance frequency are suppressed  (see centre of \cref{fig:RTC}\textbf{c}).
Although there are $N-1$ eigenstates of the Hamiltonian lying within the gap, these are purely atomic, and the symmetry of the all-to-all atom--cavity coupling prevents their excitation, rendering them completely dark. 

Upon introducing disorder, we observe the onset of a non-zero response around zero detuning, a manifestation of the increase of photon weight of the originally dark purely atomic states. A representative spectrum of $\chi_\mathrm{a}$ for $W/(2\pi)=\SI{26}{\MHz}$ is presented in \cref{fig:RTC}\textbf{d}. We observe that the fading out of the Rabi splitting occurs via a redistribution of the spectral weight from the polaritons to a wide spectrum of midgap states. For $|\Delta_{\mathrm{c}\bar{\mathrm{a}}}| \gtrsim W$, a narrow, dispersively shifted cavity resonance is restored around $\Delta_{\mathrm{pc}} = 0$ (see \cref{fig:RTC}\textbf{d}).

\begin{figure*}
\includegraphics{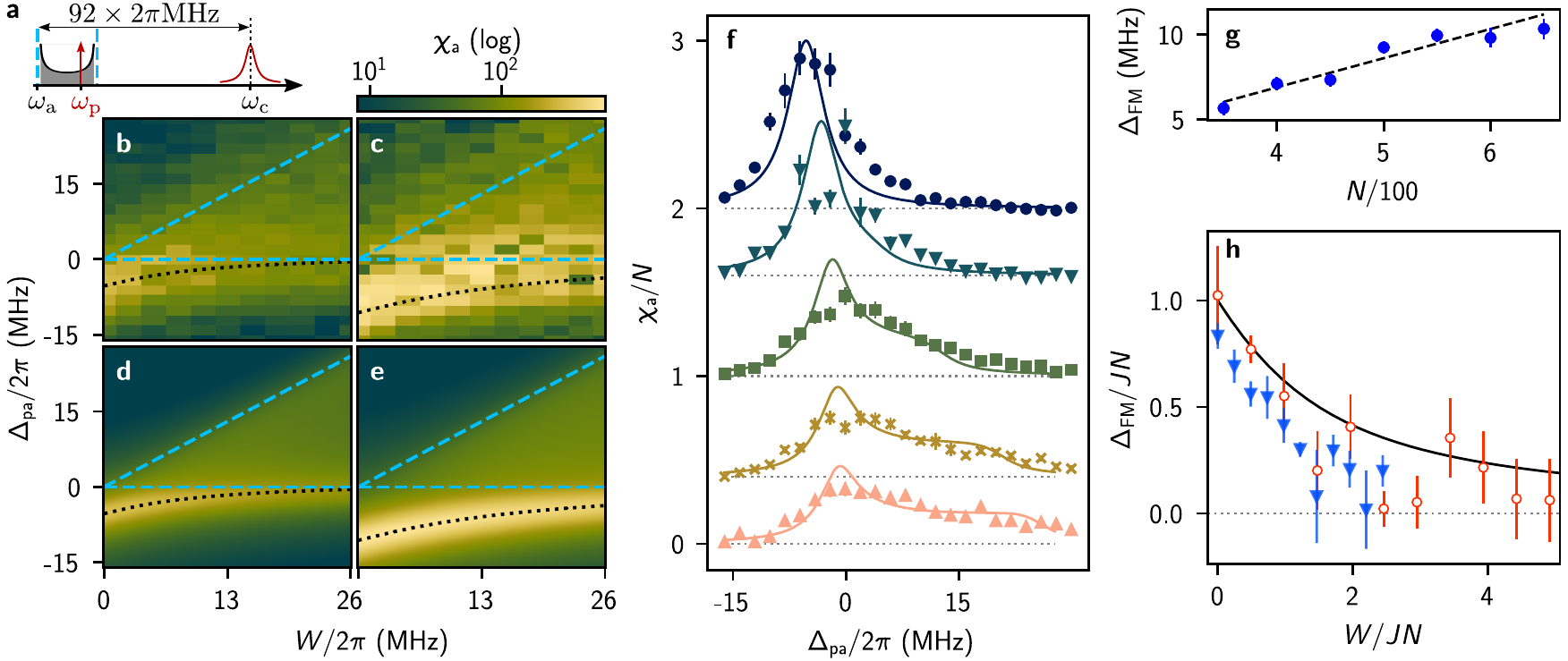}
    \caption{\label{fig:RLMG}\textbf{Response of the random LMG model.} \textbf{a}, Frequency diagram  depicting the detuning between the atomic disorder, the cavity and the weak cavity probe. \textbf{b}--\textbf{e}, Measured (\textbf{b}, \textbf{c}) and simulated (\textbf{d}, \textbf{e}) atomic susceptibility for $N = 303 \pm 63$ (\textbf{b}, \textbf{d}), and $N  = 610 \pm 57$ (\textbf{c}, \textbf{e}) atoms.
    \textbf{f}, Cuts through \textbf{b}~and~\textbf{d}, illustrating the quantitative agreement between experiment (markers) and theory (solid lines).
    Cuts show data for different disorder strengths $W/2\pi = 0.0,5.2,13.0,20.8,26.0$ $(\mathrm{MHz})$ (top to bottom), and are offset from one another according to $(26-W/2\pi )/13$. 
    \textbf{g}, Scaling of the collective ferromagnetic gap $\Delta_{\mathrm{FM}}=J N $ at zero disorder $W=0$ with mean atom number $ N $.
    \textbf{h}, Behaviour of $\Delta_{\mathrm{FM}}$ as a function of disorder strength $W$ for $ N  = 303 \pm 63$ (emtpy red circles) and $ N  = 610 \pm 57$ (blue triangles) atoms.
    To illustrate the scale invariance of the system, the axes are rescaled by the zero-disorder ferromagnetic gap size $J N $.
    Markers represent the experimental data with statistical error bars, and the lines show the theoretical results obtained by exact diagonalisation (see Methods, Sec.~\ref{s:ED_lmg}).
}
\end{figure*}

To further understand the evolution of the spectrum with disorder strength, we probe the photonic susceptibility at $\Delta_{\mathrm{c}\bar{\mathrm{a}}} = 0$ as a function of disorder strength $W$, and detuning $\Delta_{\mathrm{pc}}$.
The results are presented in \cref{fig:RTC}\textbf{g},~\textbf{i} for different mean atom numbers $N$. For weak disorder, the photonic susceptibility $\chi_{\mathrm{p}}$ confirms the presence of the two bright polaritons, and a manifold of degenerate dark states at the centre of the Rabi gap.
As the disorder becomes comparable with the collective atom--cavity coupling, $W\sim g\sqrt{N}$, we observe a smooth increase of $\chi_{\mathrm{p}}$ around $\Delta_{\mathrm{pc}}=0$, signalling the onset of a finite coupling of a grey state manifold emerging from the originally dark states. Simultaneously, the polaritons' response weakens and fades away for the largest disorder, where the spectrum consists of a resonance centred at $\Delta_{\mathrm{pc}} =0$ strongly broadened by the disorder. 

The evolution of the spectrum with disorder is driven by the fragmentation of the eigenstates, from fully delocalised bright and dark states without disorder, to randomly distributed, isolated resonances for the largest disorder. To confirm this interpretation, we compare our observation (\cref{fig:RTC}\textbf{g, i}) with theoretical calculations (\cref{fig:RTC}\textbf{h, j}) of the cavity transmission based on Green function techniques (see Methods, \cref{s:near-res-theory}).
The model takes into account the experimental distribution of the spin energies, which is correlated and non-uniform, differently from the case studied in Ref.~\cite{dubailLargeRandomArrowhead2022}. 
Nevertheless, we have verified that the eigenfunctions are multifractal in the same way (see Supplementary Information, \cref{s:multifrac}). The simulations, which take into account the measured atom number fluctuation and the effect of the thermal motion on the atom--cavity couplings, are in good agreement with the observations for the low disorder regime. For the strongest disorder, deviations appear in particular for the upper polariton, whose signal appears moderately weaker in the experiment. We attribute this to losses induced by radiation pressure from the control laser at $\SI{460}{\nm}$, affecting predominantly excited atoms with the largest admixture in the $4D_{5/2}$ manifold (see Methods). For the largest disorder strength, we do not resolve the polaritons themselves but observe a clear signal from the grey states. These results are further confirmed in \cref{fig:RTC}\textbf{k}, which presents a direct comparison of experimental and theoretical data for the photonic susceptibility $\chi_{\mathrm{p}}$ as a function of $\Delta_{{\rm pc}}$ for representative values of the disorder strength $W$. The same simulation procedure reproduces also the atomic susceptibility $\chi_{\mathrm{a}}$  measured as a function of detunings, as shown in \cref{fig:RTC}\textbf{e, f}. 

We quantitatively analyse the fading out of the polariton and the emergence of grey states by comparing the photonic susceptibility in the lower (respectively middle) parts of the spectrum shown in \cref{fig:RTC}\textbf{g}--\textbf{j}. This yields the overall photon weight of the polariton and grey states as a function of normalised disorder strength shown in \cref{fig:RTC}\textbf{l}. The crossover between the light--matter interaction dominated regime and the disorder dominated regime is manifest as spectral weight is smoothly transferred from the polariton to grey states, in qualitative agreement with the simulations.

\section{Large-detuning regime and Lipkin--Meshkov--Glick model}
\label{s:LMG}

In the central mode model investigated so far, an essential role is played by the finite admixture of the spin excitations to the delocalised photon field. For large detuning $\Delta_{\mathrm{ca}} \gg g \sqrt{N}$, the cavity field is only virtually populated, giving rise to an all-to-all interaction between the spins, thereby realising an effective Lipkin--Meshkov--Glick (LMG) model \cite{lipkinValidityManybodyApproximation1965,makhalov_2019aa,munizExploringDynamicalPhase2020} (see \cref{fig:RLMG}\textbf{a} and Methods, Sec.~\ref{s:Heff_lmg}). In the presence of a longitudinal random field, the Hamiltonian for these effective dynamics reads
\begin{equation}\label{eq:LMG_hamiltonian}
\hat{H}_\mathrm{LMG} =   \sum_{i=1}^N \epsilon_{i} \frac{\sigmah_i^z}{2} - JN \hat{S}^+ \hat{S}^-,
\end{equation}
where $J = g^2/\Delta_{\mathrm{ca}}$ is the strength of the spin-exchange interactions. 
Equation~\eqref{eq:LMG_hamiltonian} is a particular case of the class of exactly solvable Richardson--Gaudin models \cite{richardsonRestrictedClassExactEigenstates1963,gaudinDiagonalisationClasseHamiltoniens1976} that are ubiquitous in quantum many-body systems \cite{dukelskyExactlySolvableRichardsonGaudin2004}.

Similar to the central mode model, in the absence of disorder ($W = 0$), Eq.~\eqref{eq:LMG_hamiltonian} describes the dynamics of a collective spin within the Hilbert subspace of symmetric states. The non-linearity inherited from the spin--cavity coupling favours a ferromagnetic ground state, protected by a finite gap of size $J N$. A striking manifestation of ferromagnetism is the strong suppression of the zero--frequency magnetic response.

To realise the model of \cref{eq:LMG_hamiltonian}, we detune the cavity to the blue of the atomic transition by $\Delta_{\mathrm{ca}}/2\pi = \SI{92}{\MHz}$, and probe the system at a frequency $\omega_\textrm{p}$ in the vicinity of the bare atomic resonance $\omega_\textrm{a}$ (see Fig.~\ref{fig:RLMG}\textbf{a}).
In this regime, the transmission of the cavity is negligible such that $\chi_\mathrm{p} \sim 0$, and the atomic signal $\chi_\mathrm{a}(\Delta_{\mathrm{pa}})$ (see Eq.~\eqref{e:atomic_susc} of Methods for definition) directly measures the transverse spin susceptibility of the system at frequency $\Delta_{\mathrm{pa}} = \omega_\mathrm{p} - (\omega_\mathrm{a} + 2g^2/\Delta_{\mathrm{ca}})$ (see text below Eq.~\eqref{e:Heff_V} of Methods).
As shown in \cref{fig:RLMG}\textbf{b} and \textbf{c}, in the absence of disorder, the frequency dependence of $\chi_\mathrm{a}$ reveals the finite ferromagnetic gap, of magnitude $\Delta_{\mathrm{FM}}$, as well as the reduced zero--frequency susceptibility at $\Delta_{\mathrm{pa}} = 0$.
The signal is broadened by the finite decay rate of the excited atomic states, which reduces to a convolution of the response with the linewidth of the atomic transition (see Supplementary Information, \cref{s:lindblad}). 

We now investigate this model in the presence of disorder.
Similar to the central mode model, this breaks the description in terms of a collective spin, restoring the system's ability to explore the full Hilbert space.
For a given disorder strength $W$, the susceptibility (see \cref{fig:RLMG}\textbf{f}) shows an asymmetric peak, corresponding to a collectively enhanced response superimposed with a weak and broad background whose width traces the disorder strength (see dashed blue line in \cref{fig:RLMG}\textbf{b}--\textbf{e}).
This is a manifestation of the gradual fragmentation of the collective spin, as disorder renders the individual spins off-resonant with each other.
The peak is located at $-\Delta_{\mathrm{FM}}$, and we denote its amplitude by $\chi_{\mathrm{a}}^{\mathrm{FM}}$.

Tracking the location of this peak provides a measurement of the ferromagnetic gap as a function of $W$. Without disorder, this gap increases linearly with atom number, as shown in \cref{fig:RLMG}\textbf{g}.
With increasing disorder, it decreases smoothly towards zero, as shown in \cref{fig:RLMG}\textbf{h}, where, for low enough atom numbers, the gap is zero within our error bars. 
This demonstrates the competition between the infinite-range cavity-mediated interaction $J$ and spectral disorder $W$ for the dynamics of the effective model $\hat{H}_\mathrm{LMG}$. 

Our results are in very good agreement with a simulation of the response $\chi_{\mathrm{a}}$ of $\hat{H}_\mathrm{LMG}$ (see Methods, Sec.~\ref{s:ED_lmg}), over the entire parameter regime (see \cref{fig:RLMG}\textbf{b}--\textbf{f}): 
The simulated system sizes were set as the mean atom numbers $ N $ realised across all experimental runs, and the effect of the atoms' thermal motion on the atom--cavity coupling $g$ has been taken into account, as in the near-resonant case. The decrease of the ferromagnetic gap (see Fig.~\ref{fig:RLMG}\textbf{h}) indicates a drastic change of the system properties as disorder increases. However, in the thermodynamic limit the system is always ferromagnetic and no paramagnetic phase transition should occur.
Indeed, intuitively for any fixed disorder strength, increasing the number of atoms will always lead to an infinite number of close-to-resonance spins, enforcing ferromagnetism in the thermodynamic limit for an arbitrarily large disorder strength. However, for any finite number of atoms, there exists a disorder strength large enough to bring the ferromagnetic gap close to zero, by rendering each spin essentially spectrally isolated from all the others, thus crossing the system over into a paramagnet. 

More precisely, our simulations show that finite systems display a minimal gap at a disorder strength $W^\star$ suggestive of critical behaviour;  however, the value of $W^\star$ diverges with increasing atom number (see Supplementary Information, Sec.~\ref{s:SM_gapclosing} and Fig.~\ref{fig:ext_gapclosing}). 

\section{Localisation of excitations}

\begin{figure}[t]
    \centering
    \includegraphics{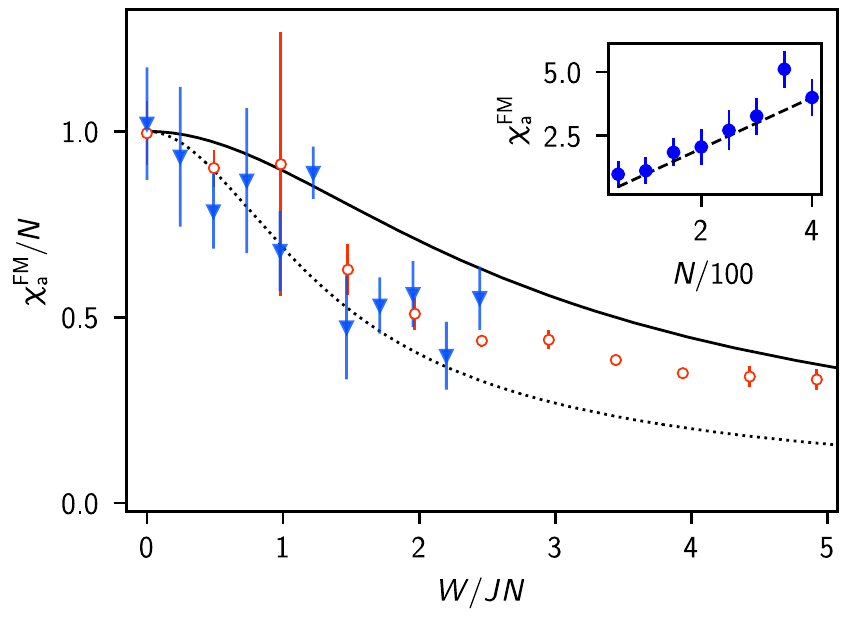}
    \caption{\label{fig:PRbound}\textbf{Participation ratio bound from atomic susceptibility.}
    Normalised atomic susceptibility $\chi_{\mathrm{a}}^{\mathrm{FM}}$, an upper bound to the participation ratio $\mathrm{PR}_1$ of the first excited state, for $N = 303 \pm 63$ (empty red circles) and $N  = 610 \pm 57$ (blue triangles) as a function of normalised disorder strength.
    The solid black line shows the corresponding simulation results for $\chi_{\mathrm{a},1}$ of \cref{e:PRbound}.
    The black dotted line is the directly simulated participation ratio of the first excited state, $\mathrm{PR}_1$.
    Inset: Maximum value of the zero-disorder atomic susceptibility as a function of atom number, showing linear scaling expected from the definition of $\chi_{\mathrm{a}}$ in \cref{e:atomic_susc}.
    }
\end{figure}

The existence and distribution of energy resonances in disordered systems is the essence of Anderson localisation. 
In our system, excitations can hop at arbitrarily large distances provided the spins are closely resonant. Disorder thus decimates the spins available for resonance by offsetting most spins from each other, but does not prevent long-distance propagation \cite{botzungDarkStateSemilocalization2020, dubailLargeRandomArrowhead2022}. 

Interestingly, while our spectroscopic probe does not yield spatially-resolved information, it does carry relevant insights about the localisation of excitations. Indeed, general arguments based on the hierarchy of \renyi entropies (see Methods, Sec.~\ref{s:PRbound}) show that a system's magnetic response may be used to bound the participation ratio of the excitations, i.e., the number of spins contributing to the wave function.
The participation ratio $\mathrm{PR}_1$ of the first excited state obeys
\begin{equation}\label{e:PRbound}
\chi_{\mathrm{a},1} \geq \mathrm{PR}_1 ,
\end{equation}
at any $W\geq 0$, where $\chi_{\mathrm{a},1}$ is the contribution of the first excited state to the atomic susceptibility when the system is probed on resonance with the transition to this state, from the global ground state (see Methods, Sec.~\ref{s:PRbound} for the proof). The bound is reached for $W=0$ where $\mathrm{PR}_1 = N$ corresponds to a wavefunction uniformly distributed over all spins, as well as in the limit $W \to \infty$ in which the excitation becomes localised on a single spin ($\mathrm{PR}_1 \to 1$). Our frequency resolved measurement thus allows us to verify the fragmentation of the system's collective excitations into ever-more localised wave-functions, consistent with the expectations for eigenstates of the central mode model \cite{botzungDarkStateSemilocalization2020, dubailLargeRandomArrowhead2022, buccheriStructureTypicalStates2011}. 

Figure \ref{fig:PRbound} shows the bound to the participation ratio deduced from our measurements, showing a decrease by more than a factor of two as disorder reaches the largest values. Upon normalisation of $\mathrm{PR}_1$ by the mean atom number $ N $, and of $W$ by the corresponding zero-disorder ferromagnetic gap $J  N $, all the data collapse onto each other and agree with simulations. 
The figure shows also the theoretically predicted value of $\mathrm{PR}_1$, which obeys the bound observed in the data. 

Similar to the ferromagnetic gap, suggestive as these findings are, they do not herald a transition from delocalised to localised. For a fixed disorder strength, increasing the number of atoms leads to an infinite number of close-to-resonance spins at arbitrary distances, preventing full localisation but leading to a semi-localised regime similar to the critical regime of the Anderson transition \cite{botzungDarkStateSemilocalization2020}.

\section{Conclusion}

Our ability to introduce controlled disorder in cavity-QED offers many timely and exciting prospects for further investigations, such as the study of Bardeen--Cooper--Schrieffer superconductivity as proposed in Ref.~\cite{lewis-swanCavityQEDQuantumSimulator2021}, where our atomic susceptibility measurements would directly map to the pairing gap. More broadly, Eq.~\eqref{eq:LMG_hamiltonian} allows the direct simulation of Richardson--Gaudin models that are relevant to a variety of many-body systems, from superconductivity in ultrasmall grains to quark physics and neutron stars. Furthermore, the capabilities demonstrated in our experiment could also be used to study the effect of inhomogeneous broadening for quantum optics applications, in particular for superradiant laser clocks \cite{bychekSuperradiantLasingInhomogeneously2021}.

While the finite lifetime of the employed excited state limits the current investigations to one excitation above the fully polarised state, higher excitations can be probed by encoding the spins in the ground state manifold and coupling them via Raman transitions \cite{Davis:2019aa} or through the use of atoms with long-lived excited states \cite{munizExploringDynamicalPhase2020}. Last, using high resolution optics and time-resolved manipulation of the control light, it will become possible to program the otherwise homogeneous long-range cavity-mediated interaction in space and time, lifting one of the most stringent restrictions for the use of cavities in quantum simulation applications. In combination with small ultra-cold samples of our Fermionic $^6$Li atoms, this will allow for the creation of random long-range interactions between Fermionic degrees of freedom, one of the building blocks for holographic quantum matter \cite{chowdhury:2021te}.

\section*{Acknowledgements}

N.S., F.O., T.C-M. and J-P.B. acknowledge funding from the Swiss National Science Foundation (grant No 184654), the Sandoz Family Foundation and EPFL.
G. P. acknowledges support from the Institut Universitaire de France (IUF) and the University of Strasbourg Institute of Advanced Studies (USIAS), the ANR via CLIMAQS. P.U., S.B. and P.H. acknowledge funding from the ERC Starting Grant StrEnQTh (project ID 804305), Provincia Autonoma di Trento, and by Q@TN, the joint lab between University of Trento, FBK-Fondazione Bruno Kessler, INFN-National Institute for Nuclear Physics and CNR-National Research Council. S.B. acknowledges CINECA for the use of HPC resources under ISCRA-C project ISSYK-2 (HP10CP8XXF).

	\bibliography{random-spin-models}

\pagebreak
\section{\label{sec:methods}Methods}

\setcounter{figure}{0} 
\renewcommand{\thefigure}{E\arabic{figure}} 
\renewcommand{\figurename}{EXTENDED DATA FIG.} 

\subsection{Experimental apparatus}
\label{sec:setup}
The core of our setup is a high-finesse optical resonator placed inside an ultra-high vacuum chamber \cite{sauerweinVibrationDampingPlatform2022}. The cavity has a finesse of \SI{59e3}{} and \SI{13e3}{} at \SI{1342}{\nm} and \SI{671}{\nm}, respectively. The cavity is \SI{25.9}{\mm} long, \SI{103}{\micro\meter} shorter than concentric, giving us a single-atom single-photon cooperativity of $\eta = 6.4$. The \SI{1342}{\nm} light is used for frequency stabilisation and dipole trapping and the \SI{671}{\nm} light allows for resonant coupling to the D$2$ transition of Lithium.

We use in total two lasers, a \SI{1342}{\nm} diode laser (main laser) that is Raman-fibre amplified and then frequency-doubled to generate light at \SI{671}{\nm}, and a laser diode emitting at \SI{460}{\nm} (light-shifting laser). The main laser is used for the magneto-optical trap (MOT), absorption imaging, cavity probing, and trapping of the atoms in a cavity-enhanced optical dipole trap. It is stabilised to our cavity on the TEM$_{04}$ mode at \SI{1342}{\nm}. The length of the cavity itself can be controlled using piezoelectric actuators under the mirrors. We can stabilise the detuning between the D$2$ transition of Lithium and the resonance frequency of our cavity in a large frequency range ($> \SI{1}{\GHz}$), by using a sideband of the \SI{671}{\nm} beam sent to a saturated absorption spectroscopy cell.
A feed-forward scheme acting on both the cavity and the laser allows us to rapidly vary the cavity--atom detuning within the experimental sequence (max.\ slew rate of \SI{0.1}{\GHz \per \ms}) while holding the atoms in the cavity dipole trap. The light-shifting laser is stabilised using a commercial wavemeter.

\subsection{Atoms preparation}\label{s:atomprep}
We prepare an atomic cloud with a target atom number and size using a combination of laser cooling, spatial selection, and cavity-assisted feedback techniques. We start by loading the atoms from a MOT directly into the intra-cavity standing-wave dipole trap, with a temperature of about \SI{200}{\micro \kelvin} and trap frequencies of $\omega_{\perp}/2\pi = \SI{22}{\kHz}$ and $\omega_{\parallel}/2\pi = \SI{1.4}{\MHz}$ in the transverse and longitudinal directions, respectively.

At this point, the cavity resonance frequency is set \SI{1}{\GHz} red detuned with respect to the D$2$ transition. We then start an optical molasses phase using the MOT beams, while probing the cavity using light detuned by a fixed amount with respect to the resonance of the empty cavity. The dispersive shift of the cavity is reduced as atoms are lost during the molasses, until the probe becomes resonant with the cavity, leading to an increased transmitted photon flux detected by a single photon counter.
The molasses is stopped when the target atom number set by the predefined dispersive shift is reached and the sequence can continue.
When turning off the optical molasses beams, we make sure that all atoms are optically depumped into the ${2S_{1/2}^{F=1/2}}$ manifold.
 
At this point of the sequence, the atomic cloud measures a length of \SI{0.5}{\mm}, populating about $750$ pancakes, each containing between $0.4$ and $4$ atoms on average. We empty all but the central $180$ the sites using radiation pressure, by imaging an opaque mask on the centre of the cloud with a laser resonant on the D$2$ transitions, as presented in \cref{fig:extended_blue}\textbf{a}. We then shift within \SI{30}{\ms} the cavity on resonance with the ${2S_{1/2}^{F = 3/2}}$--$2P_{3/2}$ transition, leading to a detuning of \SI{228}{\MHz} (hyperfine splitting of $^6$Li) with respect to the ${2S_{1/2}^{F = 1/2}}$--$2P_{3/2}$ transition resonant with the atoms.

We then perform a fast cavity transmission spectroscopy, by sweeping a weak probe over the cavity resonance. The dispersive shift of the cavity is used to extract the initial number of atoms in the $F=1/2$ state. A similar sweep is performed after the interrogation of the disordered system. Together, they allow for the characterisation of probe-induced atom losses.

\subsection{Implementation of the disorder}
\begin{figure}
	\centering
	\includegraphics[width = 0.47\textwidth]{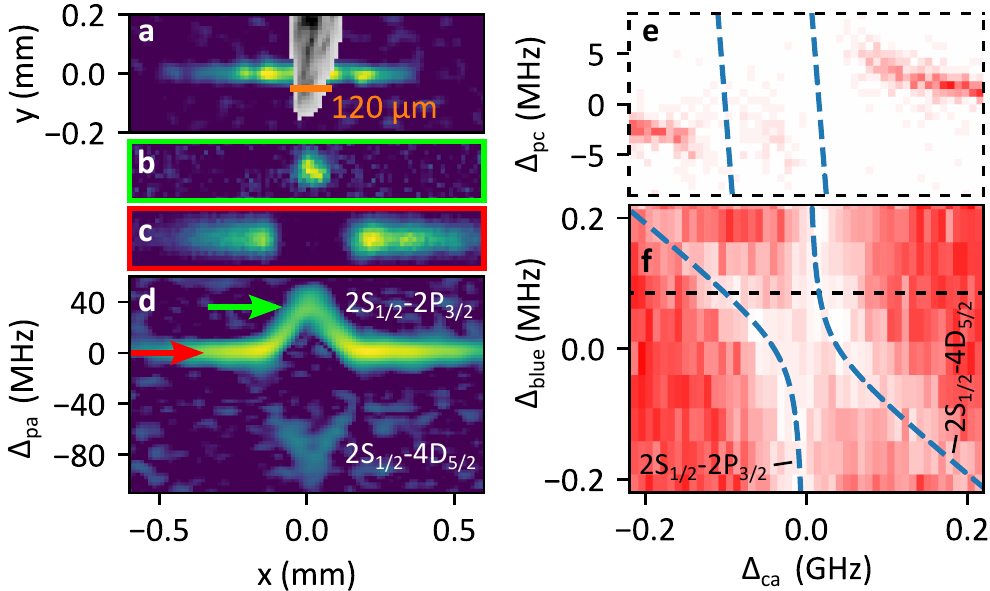}
	\caption{\label{fig:extended_blue}\textbf{Characterisation of light-shift of excited state.} \textbf{a}, Absorption picture of an atomic cloud in the cavity dipole trap. The grey needle in the middle is used to shield central atoms from the absorption light, making it possible to empty the outer pancake traps using the radiation pressure force of the absorption imaging. \textbf{b} and \textbf{c}, Absorption image of the cloud at maximal shift (green arrow in \textbf{d}) and zero shift (red arrow in \textbf{d}). \textbf{d}, Spatially-resolved absorption spectroscopy of the D$2$ transition in the presence of a single lattice beam (Gaussian beam centred on $x = \SI{0}{\mm}$ with waist of \SI{120}{\micro \m}). Each horizontal line of plot \textbf{d} is the $y$-integrated optical density. The light shifting laser was \SI{50}{\MHz} blue detuned from the $2P_{3/2}$--$4D_{5/2}$ transition. We can observe both resonances of the Autler--Townes doublet, the strong light-shifted single photon transition (top) and the faint two-photon transition (bottom). \textbf{e}, Cavity transmission spectroscopy for cloud illuminated with homogenous light-shifting beam. The dashed blue lines indicate the expected resonance frequencies of the Autler--Townes doublet. \textbf{f}, $2$D map of cavity transmission spectra for different detunings of the light-shifting laser from the $2P_{3/2}$--$4D_{5/2}$ transition. Each horizontal line is a cavity transmission spectrum. In the colorbar, red indicates a large number of photons transmitted through the cavity. The black horizontal dashed line marks the configuration of panel \textbf{e}.}
\end{figure}

We encode the two-level system using the ${2S_{1/2}^{F=1/2}}$ ($\ket{g}$) and $2P_{3/2}$ ($\ket{e}$) states of our $^6$Li atoms. The transition frequency of the atoms can be tuned by light-shifting the excited state $\ket{e}$. In particular, this is achieved by dressing the $2P_{3/2}$ state with the higher-lying $4D_{5/2}$ manifold using a control laser at \SI{460}{\nm} detuned from resonance (see \cref{fig:intro}\textbf{c}) by $\Delta_\mathrm{blue}$. We first calibrated the light-shift of the excited state---due to a single Gaussian beam of waist \SI{120}{\micro \m}, with $\Delta_\mathrm{blue}=$\SI{50}{\MHz}---by performing an absorption spectroscopy of the D$2$ transition similar to \cite{brantutLightshiftTomographyOpticaldipole2008}. Taking absorption images of the cloud at different imaging frequencies, we reconstructed the spacial distribution of the light-shift of a single lattice beam as presented in \cref{fig:extended_blue}\textbf{b}. We performed this spectroscopy both in-situ and after releasing the atoms from the cavity dipole trap, allowing us to measure the trap-related shift of the $2P_{3/2}$--$4D_{5/2}$ transition to be \SI{90}{\MHz}. 

Furthermore, we characterised the dependence of the cavity transmission spectrum on the detuning of the light shifting laser, showing an avoided crossing for both states of the Autler--Townes doublet, in particular the light-shifted single-photon $2S_{1/2}$--$2P_{3/2}$ transition and the two-photon transition $2S_{1/2}$--$4D_{5/2}$ (see \cref{fig:extended_blue}\textbf{c}). We observed increased atom losses for small detunings of the light-shifting laser, pointing towards radiation pressure induced atom losses, occurring when atoms are promoted to the $2P_{3/2}$ state during the spectroscopic measurements. We minimised this effect by choosing the maximal detuning (\SI{400}{\MHz} blue detuned from the $2P_{3/2}$--$4D_{5/2}$ transition), allowing us to go up to $W = \SI{26}{\MHz}$ for the maximal available power of the laser of \SI{7.3}{\mW} per lattice beam.

The light-shifting lattice beams are linearly polarised perpendicular to the cavity axis, and set the direction of the quantisation axis. We then probe the cavity using $\pi$-polarised light, to avoid any vector light-shift effect of the light-shifting beam. Because atoms reside in the $F=1/2$ hyperfine manifold, the $\pi$ transition used for cavity interrogation is free of tensor light-shift effects. As a result, even though our sample comprises an incoherent mixture of the two magnetic sublevels of the $F=1/2$ manifold, the two components experience a strictly identical light shift and probe beam, contributing equally to the signal without further broadening effects. Cross-optical pumping between the two does not deteriorate the signal in the linear response regime explored in this work. 

\subsection{Interrogation}
\label{s:measure_atom_excitations} 
Once the preparation phase is completed, we tune the cavity to the desired length and illuminate the cloud with the light-shifting lattice. We send a cavity probe pulse with a duration of \SI{5}{\micro \second} or \SI{60}{\micro \second} for the measurements presented in \cref{fig:RTC} and \cref{fig:RLMG}, respectively. During this measurement, we monitor the photons leaking out of the cavity using a single photon counter, to infer the optical response. At zero magnetic field, the transition between $\ket{g}$ and $\ket{e}$ is not closed, and an atom in state $2P_{3/2}$ can decay into the $F=3/2$ ground state manifold, denoted as an auxiliary state $\ket{a}$. This state is not coupled to the cavity field, thanks to the large hyperfine splitting. Since the decay can only happen from state $\ket{e}$, the population accumulated in the $F=3/2$ state is directly proportional to the excited state population $\langle \hat{S}_z\rangle$ integrated over the probe pulse duration. 

\begin{figure}
	\centering
	\includegraphics[width = 0.47\textwidth]{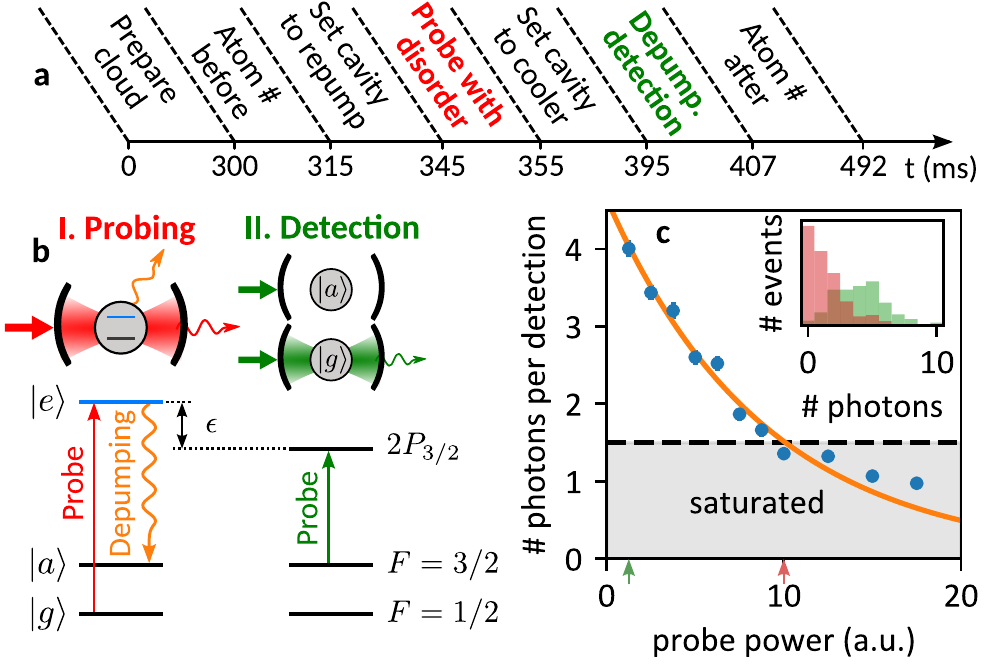}
	\caption{\label{fig:extended_seq} \textbf{a}, Timeline of the experimental sequence. The core elements are the interrogation of the disordered cavity--atom system (red) and the subsequent detection of depumped atoms in the $\ket{a}$($F = 3/2$) state. \textbf{b}, Illustration of the probe configurations during interrogation and depumping detection. \textbf{I}, When probing the disordered system, the excited state of the atoms is dressed with the light-shifting laser, indicated by the blue colour of the level and the shift $\epsilon$. The photons entering the cavity from the probe have two decay channels. Either they leak out of the cavity on the other side (red wiggled arrow) where they will be detected by a single photon counter, or they can be lost by free-space spontaneous emission of an atomic excitation (orange wiggled line). At zero-magnetic field, the transition from $\ket{g}$ to $\ket{e}$ is not closed, therefore spontaneous emission events can depump the atom into the $\ket{a}$ state. \textbf{II}, These depumped atoms can be detected by measuring the cavity transmission. If the cavity is on resonance with the $\ket{a}$--$\ket{e}$ transition and only a single atoms is in state $\ket{a}$ the transmission gets strongly suppressed. \textbf{c}, Calibration of depumping signal. The mean number of transmitted photons during the depumping detection is plotted against the probe power during the interrogation of the disordered system. The error bars of the data represent statistical fluctuations. The orange line shows a fit of the theoretically expected relation [see \cref{e:me_popA}]. The inset shows a histogram of detected photons for low (green) and high (red) probe powers (see arrows on $x$-axis for configuration).}
\end{figure}

The population of the $F=3/2$ state is measured after the interrogation of the disordered system using a cavity transmission spectroscopy, with the cavity tuned on resonance with the $F=3/2$ to $2P_{3/2}$ transition (see figure \cref{fig:extended_seq}\textbf{b} right). In this configuration, the cavity transmission is suppressed by $1 / (\eta + 1)$ in the presence of a single atom in the $F=3/2$ state, yielding a single-atom level sensitivity for the detection of atomic response.

In practice, we implement the detection by sweeping the frequency of the on-axis probe over the cavity resonance yielding an average photon count of $4$ photons per sweep for the empty $\ket{a}$ manifold (see green histogram in inset of \cref{fig:extended_seq}\textbf{c}). The frequency sweep is essential since it removes the systematic effects of dispersive shifts on the depumping detection resulting from the presence of atoms in the $\ket{g}$ state. \cref{fig:extended_seq}\textbf{c} shows the dependence of the number of transmitted photons on the laser power during the interrogation, showing the expected exponential trend (see \cref{s:meas_atomic_sus}), allowing for the characterisation of the atomic susceptibility. At large probe powers, we observe a deviation from the exponential model that is due to saturation effects and atom losses. The data presented in this work was measured at different probe powers, and measurements with an average photon count below $1.5$ photons per sweep were neglected (see dashed line in \cref{fig:extended_seq}\textbf{c}), ensuring that no additional broadening of the resonances is introduced.

\subsection{Susceptibility in the near-resonant regime}
\label{s:near-res-theory}

In this section, we provide some details on the calculations of the susceptibility in the near-resonant regime, whose results are presented in \cref{s:near-resonant} of the main text.

In our calculations, we account for fluctuations both in the atom number $N$ and in the atom--cavity couplings $g$. Specifically, we average the susceptibility over $1000$ different values of $N$ randomly sampled from a normal distribution. The mean and standard deviation of the $N$ distribution have been determined by fitting the experimental data at $W=0$ and they are reported in
the caption of \cref{fig:RTC}.
For each value of $N$, we consider a generalised version of the TC Hamiltonian \cref{eq:Hamiltonian}, namely
\begin{equation}\label{eq:HamiltonianRand}
\hat{H}_\mathrm{TCr} =  \Delta_{\mathrm{ca}} \ah^{\dagger}\ah+  \sum_{i=1}^N g_i \left(\sigmah_i^+ \, \ah + \sigmah_i^- \, \ah^{\dagger} \right) + \sum_{i=1}^N \epsilon_{i} \, \frac{\sigmah_i^{z}}{2}~,
\end{equation}
where the couplings $g_i$ are randomly generated accounting for the finite temperature of the atoms and the polarisation of the probe light. We also account for the fact that the $N$ atoms are randomly distributed across $\approx100$ pancakes: to this end, we randomly select $N$ site energies among the set $\epsilon_{i} \in \{ \frac{W}{2}\cos (2\pi\lambda_lj/\lambda_s)~,~j=1,\dots,100\}$, where $\lambda_l=671$~nm is the lattice wavelength and $\lambda_s=1040$~nm is the light-shift wavelength.

For each value of $N$, following Refs.~\cite{dubailLargeRandomArrowhead2022,leslieTransmissionSpectrumOptical2004}, we employ a Green function formalism in the linear response regime. 
In such a situation, the cavity susceptibility (cavity transmission) at a given probe--cavity detuning $\Delta_{\rm pc}$ is
\begin{equation}\label{chip_definition_onres}
\chi_{\rm p}(\Delta_{\mathrm{pc}})\propto-{\rm Im}\left(\bra{G} \ah  \frac{1}{\Delta_{\mathrm{pc}}- \hat{\cal H}} \ah^\dag \ket{G}\right)\,,
\end{equation}
where $\ket{G}$ is the ground state. In \cref{chip_definition_onres}, we introduced the non-Hermitian Hamiltonian
\begin{equation}
\hat{\cal H} = \hat{H}_{\rm TCr} -i\frac{\Gamma}{2}\sum_{i=1}^N \sigmah_i^+ \sigmah_i^- -i\frac{\kappa}{2}\ah^\dag \ah \,,
\end{equation}
which includes the generalised TC Hamiltonian~\cref{eq:HamiltonianRand} and two terms describing cavity losses and atom decay, respectively. Similarly, the atomic susceptibility is computed by summing the transition probabilities to all the atomic states $\sigmah_i^+\ket{G}$, namely
\begin{equation}\label{chia_definition_onres}
\chi_{\rm a}(\Delta_{\mathrm{pc}})\propto\sum_{i=1}^N\left|\bra{G} \sigmah_i^-  \frac{1}{\Delta_{\mathrm{pc}}- \hat{\cal H}} \ah^\dag \ket{G}\right|^2~.
\end{equation}

\subsection{Measurement of atomic susceptibility}
\label{s:meas_atomic_sus}

We now show that the atomic susceptibility $\dsa$, discussed in Sec.~\ref{s:LMG} of the main text, can be extracted from measurements of the atomic population $P_\mathrm{A}(t)$ of the auxiliary state $\ket{a}$ at a given point in time $t$.
Intuitively, it is plausible that $\dsa$ and $P_\mathrm{A}(t)$ should be connected:
On the one hand, $\dsa$ is simply a rescaling of the absorptive part of the dynamic susceptibility $\ds$ [see Eq.~\eqref{e:chi_full}--\eqref{e:atomic_susc}] of the effective model described by Eq.~\eqref{e:Heff}, and thus quantifies the time-averaged energy absorbed by this system when subjected to a perturbation at frequency $\Delta_{\mathrm{pa}}$.
On the other hand, the system can absorb energy from the probe beam \emph{only} via coherent excitations of the atomic population from state $\ket{g}$ (${2S_{1/2}^{F = 1/2}}$) to $\ket{e}$ ($2P_{3/2}$). The population of state $\ket{a}$ (${2S_{1/2}^{F = 3/2}}$) can then change \emph{only} via spontaneous decay from state $\ket{e}$ at a rate $\Gamma_\mathrm{a}$.
Therefore, detecting $P_\mathrm{A}(t) > 0$ implies that the system has absorbed energy via atomic excitations.
Furthermore, the probability to excite the system into a collective state containing an atomic excitation upon probing is maximised when the probe frequency $\Delta_{\mathrm{pa}}$ is resonant with transitions from the system's collective ground state.
It follows that the total atomic population $P_\mathrm{A}(t_\mathrm{meas})$ found in state $\ket{a}$ after the interrogation time is a measure of how \emph{susceptible} the effective model was to excitations introduced by the probe at frequency $\Delta_{\mathrm{pa}}$.

We give the above intuition an analytic foundation by modelling the experimental sequence of Secs.~\ref{s:atomprep}--\ref{s:measure_atom_excitations} via a Lindblad master equation (see  Supplementary Information \cref{s:lindblad} for details of the derivation).
We derive an equation of motion for $P_\mathrm{A}(t)$ in terms of $\dsa$ using that the probe beam's amplitude $\Omega_{\mathrm{p}}$ is much weaker than the natural linewidth $\Gamma$  of $^6$Li, i.e., that atomic excitations decay at a rate much faster than the rate at which they are introduced by the probe beam, $\abs{\Omega_{\mathrm{p}}} \ll \Gamma$.
This yields the relation
\begin{equation}\label{e:me_popA}
P_\mathrm{A}(t_\mathrm{meas}) =  1 - \exp( - \frac{ \Gamma_\mathrm{a}}{(\Gamma/2)^2}  \left\lvert \frac{g \Omega_{\mathrm{p}}}{\Delta_{\mathrm{ca}}} \right\rvert^2 \chi_{\mathrm{a}}(\Delta_{\mathrm{pa}})   t_\mathrm{meas})  ,
\end{equation}
evaluated here at the measurement time $t=t_\mathrm{meas}$.
This result confirms the monotonic relation between $P_\mathrm{A}(t)$ and $\dsa$.
It is obtained with respect to the experiment's initial conditions $p_G(0)=1, P_\mathrm{A}(0)=0$, and is valid for times ${t \gg (\Gamma/2)^{-1}}$.

The saturation of $P_\mathrm{A}(t_\mathrm{meas})$ as a function of the probe power $\abs{\Omega_{\mathrm{p}}}^2$, as illustrated by the data of \cref{fig:extended_seq}\textbf{c}, is captured by Eq.~\eqref{e:me_popA}.
Further, for a given probe power, the saturation rate is maximal at those probe frequencies $\Delta_{\mathrm{pa}}$ at which $\dsa$ is largest:
Since population transfer from $\ket{G}$ to a state $\ket{m}$ of the single-excitation manifold (SEM, see Sec.~\ref{s:Heff_lmg}) is maximised when the probe frequency is resonant with the transition frequency $E_{mG}$ [i.e., resonant with a frequency at which the system is most \emph{susceptible} to perturbations, as quantified by $\dsa$], the concomitant accumulation of population in the auxiliary state is also maximised.
Conversely, for a fixed measurement time $t_\mathrm{meas}$, saturation of the signal $P_\mathrm{A}(t_\mathrm{meas})$ can be suppressed by reducing the probe's power.
This is crucial for the precision of the experimental data presented in Fig.~\ref{fig:RLMG}, as elaborated in Sec.~\ref{s:measure_atom_excitations}, where the experimental technique for measuring $P_\mathrm{A}(t_\mathrm{meas})$ is discussed.

\subsection{Effective model and \texorpdfstring{$\chi_\mathrm{a}$}{χₐ} in the large-detuning regime}
\label{s:Heff_lmg}

In this section, we demonstrate that the dynamics of our system are described by the effective Hamiltonian of Eq.~\eqref{eq:LMG_hamiltonian} when the cavity is tuned far into the dispersive regime, such that $\Delta_{\mathrm{ca}} \equiv \omega_\mathrm{c} - \omega_\mathrm{a}$ is the dominant energy scale.

Our starting point is the disordered Tavis--Cummings Hamiltonian $\hat{H}_\mathrm{TC} $, which is expressed in Eq.~\eqref{eq:Hamiltonian} relative to the rotating frame (RF) of the \emph{bare} atomic resonance frequency $\omega_\mathrm{a}$.
Within this RF, the probe beam is described by the perturbation ${\hat{V}(t) = \Omega_\mathrm{p} e^{-i (\omega_\mathrm{p} - \omega_\mathrm{a}) t} \hat{a}^\dagger  + \mathrm{h.c.} }$, with probe-laser and Rabi frequency $\omega_\mathrm{p}$ and $\Omega_{\mathrm{p}}$, respectively.

The total Hamiltonian is thus $\hat{H}_\mathrm{TC} + \hat{V}(t)$, and thus the equation of motion of the (Heisenberg picture) photonic operator $\hat{a}(t)$ is
\begin{equation}
\partial_t \hat{a}(t) = -i\left[\hat{a}(t), \hat{H}_\mathrm{TC} + \hat{V}(t) \right] - (\kappa/2)\hat{a}(t),
\end{equation}
where the last term takes into account cavity losses.
Using that these are sub-dominant, i.e., $\Delta_{\mathrm{ca}} \gg \kappa$ (see Sec.~\ref{s:model}~and\ref{s:LMG} of the main text), the cavity mode adiabatically follows the evolution of the spin degrees of freedom as
\begin{equation}
\hat{a}(t) = -\frac{ g \sqrt{N} \hat{S}^- + \Omega_{\mathrm{p}} e^{-i (\omega_\mathrm{p} - \omega_\mathrm{a}) t} }{ \Delta_{\mathrm{ca}} } .
\end{equation}
Substituting this expression into $\hat{H}_\mathrm{TC} + \hat{V}(t)$ eliminates the cavity mode, and one obtains (up to an irrelevant constant term) the effective spin Hamiltonian
\begin{eqnarray}
&&\hat{H}(t) = \hat{H}_\mathrm{LMG} - \hat{\mathcal{V}}(t)\,,\quad \text{ where} \label{e:Heff} \\
&&\hat{\mathcal{V}}(t) = \frac{g \sqrt{N} }{\Delta_{\mathrm{ca}}} \left( \Omega_{\mathrm{p}}  e^{-i\Delta_{\mathrm{pa}} t} \hat{S}^+ + \mathrm{h.c.} \right) , \label{e:Heff_V}
\end{eqnarray}
with $\Delta_{\mathrm{pa}} \equiv \omega_\mathrm{p} - (\omega_\mathrm{a} + 2g^2 /\Delta_{\mathrm{ca}})$, and $\hat{H}_\mathrm{LMG}$ as given by Eq.~\eqref{eq:LMG_hamiltonian} of the main text.
We note that the above equations are obtained after performing an additional RF transformation, which serves only to remove an otherwise constant contribution $(2g^2 N /\Delta_{\mathrm{ca}})\hat{S}^z$ to Eq.~\eqref{eq:LMG_hamiltonian}.

Having obtained the above effective model, we now derive the form of the atomic susceptibility $\dsa$ in the dispersive regime.
In particular, $\dsa$ is obtained from the absorptive part $\ds$ of the dynamic susceptibility of the effective model $\hat{H}_\mathrm{LMG}$ of Eq.~\eqref{eq:LMG_hamiltonian}, when the latter is initialised in its ground state $\ket{G} \equiv \bigotimes_{i=1}^N \ket{g}_i$, and subsequently subjected to the probe via the interaction $\hat{\mathcal{V}}(t)$ of Eq.~\eqref{e:Heff_V}.
Studying the dynamic susceptibility is motivated by the fact that the probe beam is weak $\abs{\Omega_{\mathrm{p}}} \ll \Gamma$, so that one may treat $\hat{\mathcal{V}}(t)$ as a perturbation within the regime of linear response \cite{jensenMackintoshRareEarthMagnetism1991}.
In particular, $\abs{\Omega_{\mathrm{p}}} \ll \Gamma$ implies that atomic excitations decay much faster than the rate at which they are introduced into the system, so that one may study the limit in which there is at most a single excitation present in the system.
That is, one need only consider the eigenstates $\ket{G}$ and  $\lbrace \ket{m} \rbrace_{m=1}^{N}$, where the latter set of states forms the single-excitation manifold (SEM) of $\hat{H}_\mathrm{LMG}$.
We denote the respective eigenenergies as $\mathcal{E}_G, \mathcal{E}_m $, and the spectral gaps as ${E_{mG} \equiv \mathcal{E}_m - \mathcal{E}_G}$, for $m=1,\ldots, N$.
With respect to this basis, we then have
\begin{equation}\label{e:chi_full}
\ds = \pi \sum_{m \in \mathrm{SEM}} \abs{  \mel{m}{ \frac{g \sqrt{N} \Omega_{\mathrm{p}}}{ \Delta_{\mathrm{ca}} }  \hat{S}^+ }{G} }^2 \delta( \Delta_{\mathrm{pa}} - E_{mG}) .
\end{equation}
In what follows, we approximate the Dirac-delta functions in Eq.~\eqref{e:chi_full} as Lorentzian responses
\begin{equation}\label{e:lorenzian_response}
\delta_\gamma(\omega) \equiv  \frac{\gamma / \pi}{ \gamma^2 + \omega^2} \text{ such that } \delta(\omega) = \lim_{\gamma \to 0} \delta_\gamma(\omega) .
\end{equation}
Here, $\gamma$ is the line-width of the (normalised) resonance, which according to the Wiener--Khintchine theorem \cite{wienerGeneralizedHarmonicAnalysis1930, khintchineKorrelationstheorieStationaerenStochastischen1934} 
corresponds to a finite experimental measurement time $1/\gamma$.
The dimensionless atomic susceptibility $\dsa$, presented in Fig.~\ref{fig:RLMG}, is then finally obtained from $\ds$ as
\begin{equation}\label{e:atomic_susc}
\begin{split}
\dsa =& \gamma \left\lvert \frac{\Delta_{\mathrm{ca}} }{g \Omega_{\mathrm{p}}} \right\rvert^2 \ds \\
=& \sum_{m \in \mathrm{SEM}} N \abs{  \mel{m}{ \hat{S}^+ }{G} }^2 \frac{\gamma^2}{ \gamma^2 + (\Delta_{\mathrm{pa}} - E_{mG})^2} \\
\equiv& \sum_{m \in \mathrm{SEM} } \chi_{\mathrm{a},m}(\Delta_{\mathrm{pa}}) .
\end{split}
\end{equation}
At zero disorder, only the first excited state $\ket{m=1} = \hat{S}^+ \ket{G}$ contributes, such that on resonance $ { \chi_{\mathrm{a}}(\Delta_{\mathrm{pa}} = E_{1 G}) = \chi_{\mathrm{a},1}(E_{1 G}) = N }$ (see Fig.~\ref{fig:PRbound} and inset).

\subsection{Participation ratio and its relation to susceptibility}
\label{s:PRbound}

Here, we prove the relation, given in Inequality~\eqref{e:PRbound}, between the atomic susceptibility and the participation ratio (PR).
The participation ratio quantifies the extent to which a given state is (de)localised over a basis of interest.
In our context, we wish to study the (de)localisation of a SEM eigenstate $\ket{m} \equiv \sum_{i=1}^N c_{mi} \sigmah_i^+ \ket{G}$ of the LMG Hamiltonian of Eq.~\eqref{eq:LMG_hamiltonian} over the spins $i$ of the system. This is quantified by the participation ratio 
\begin{equation}\label{e:PRm}
\mathrm{PR}_m = \left( \sum_{i=1}^N \lvert c_{mi} \rvert^4 \right)^{-1} \in [1,N] ,
\end{equation}
of which the limiting values $1$ and $N$ are respectively obtained at $c_{mi} = \delta_{i,i^*}$ (full localisation at some site $i^*$, achieved at $W\to\infty$), and $c_{mi} = \frac{1}{\sqrt{N}}, \, \forall \, i$ (full delocalisation over all $N$ sites, achieved at $W\to 0$).

Our proof of Inequality~\eqref{e:PRbound} relies on the identification of $\mathrm{PR}_1$ and $\chi_{\mathrm{a},1}(E_{1G})$ [see Eq.~\eqref{e:chi_m}] as monotonic functions of different \renyi entropies ${ H_{\alpha}(\Vec{p}) = \frac{1}{1-\alpha}\log(\sum_i p_{i}^\alpha) }$, and then exploiting the hierarchy $H_{\alpha_1}(\Vec{p}) \geq H_{\alpha_2}(\Vec{p})$ for any real numbers ${ \alpha_2 \geq \alpha_1 \geq 0 }$ \cite{beckThermodynamicsChaoticSystems1993}. 
To this end we note that:
(i) On resonance ${\Delta_{\mathrm{pa}}=E_{mG}}$, the $m$th summand of the atomic susceptibility defined in Eq.~\eqref{e:atomic_susc} reduces to
\begin{equation}\label{e:chi_m}
\qquad \,\chi_{\mathrm{a},m}(E_{mG}) =  \abs{ \sum_{i=1}^N c_{mi} }^2 \in [1,N] ,
\end{equation} whose limiting values are obtained with the same distributions of $c_{mi}$ as for $\mathrm{PR}_m$ (see Eq.~\eqref{e:PRm} and text thereafter).
(ii)  Using Perron--Frobenius theory \cite{meyer} one can show that the lowest SEM eigenstate $\ket{m=1}$ of $\hat{H}_{\mathrm{LMG}}$ [as defined in Eq.~\eqref{eq:LMG_hamiltonian}] satisfies $c_{1i} \geq 0, \forall i=1,\ldots, N$.
Hence, one has that $ c_{1i} = + \sqrt{p_{1i}}$, where $p_{mi} \equiv \abs{c_{mi}}^2$ are the probabilities associated to the amplitudes $c_{mi}$.

Now, for the identification with \renyi entropies, we expand both sides of Inequality~\eqref{e:PRbound} and employ point (ii). This yields
\begin{equation}
\begin{split}
&\chi_{\mathrm{a},1}(E_{1G}) =  \abs{ \sum_{i} c_{1i} }^2 = \left( \sum_{i} p_{1i}^{\alpha_1} \right)^{ \frac{1}{1-{\alpha_1} } } = \exp(H_{\alpha_1}(\Vec{p_1})), \\
&\mathrm{PR}_1 = \left( \sum_{i=1}^N \lvert c_{1i} \rvert^4 \right)^{-1} = \left( \sum_{i} p_{1i}^{\alpha_2 } \right)^{ \frac{1}{1-{\alpha_2} } } = \exp(H_{\alpha_2}(\Vec{p_1})),
\end{split}
\end{equation}
where $\alpha_1=1/2$, $\alpha_2=2$, and ${\Vec{p}_1 \equiv \left(p_{11},p_{12}, \ldots, p_{1N} \right)}$. Since $\exp(x)$ is monotonic, the hierarchy of \renyi entropies is preserved, and thus
\begin{equation}
\exp(H_{\alpha_1}(\Vec{p}_1)) \geq \exp(H_{\alpha_2}(\Vec{p}_1)) \text{ for } \alpha_1=1/2 \text{ and } \alpha_2 = 2.
\end{equation}
This concludes the proof.

We briefly comment on how the participation ratio bound of Inequality~\eqref{e:PRbound} may be measured:
Due to the finite atomic linewidth (see Sec.~\ref{s:meas_atomic_sus}), extracting \emph{only} the $m=1$st summand of the atomic susceptibility is not feasible, as nearby resonances will add to the measured signal.
What can be feasibly extracted is the amplitude $\chi_{\mathrm{a}}^{\mathrm{FM}}$ of the full susceptibility of Eq.~\eqref{e:atomic_susc}, which satisfies ${ \chi_{\mathrm{a}}^{\mathrm{FM}} \equiv \chi_\mathrm{a}(E_{1G}) \geq \chi_{\mathrm{a},1}(E_{1G}) }$, by definition.
This is the data given in Fig.~\ref{fig:PRbound}.

In closing, we note that Inequality~\eqref{e:PRbound} [as well as its looser form in terms of the full $\chi_{\mathrm{a}}(E_{1G})$] becomes an equality in both limits of $\ket{m=1}$ being fully (de)localised.
This too follows from the above expression in terms of \renyi entropies: For all $\alpha \geq 0$, ${H_{\alpha}(\Vec{p})=\log(N)}$ if $p_i = 1/N, \, \forall i=1,\ldots,N$ (maximal uncertainty), and $H_{\alpha}(\Vec{p})=0$ if $p_i = \delta_{i,i^*}$ for some $i^*=1,\ldots,N$ (maximal certainty).

The above discussion exemplifies that the participation ratio is an entropic measure, quantifying the degree of (un)certainty [(de)localisation]---obtained from some state's expansion coefficients---as to its spread over a chosen set of degrees of freedom (basis). In fact, for any basis $\lbrace \ket{i} \rbrace$, the generalised \emph{inverse} participation ratio $\mathrm{IPR}_q(\ket{\psi})$ is related to \renyi entropies via its multifractal dimension $D_q$:
Combining Eqs.~\eqref{e:genIPR}~and~\eqref{e:Dq}, one has $D_q = \frac{d}{1-q} \frac{ \log( \mathrm{IPR}_q( \ket{\psi} ) ) }{ \log(N) } $.
As in the above discussion, this can be expressed in terms of \renyi entropies as $D_q = d \frac{H_{\alpha=q}(\Vec{p})}{\log(N)}$, where $\Vec{p} = (\abs{ \braket{i=1}{\psi} }^2, \ldots , \abs{ \braket{i=N}{\psi} }^2)$.
This relation exemplifies the intimate link between entropy and quantifiers of a state's (de)localisation properties, and has as an immediate consequence that $D_q$ decays monotonically with $q \geq 0$.

\subsection{Numeric simulation of the large-detuning regime}
\label{s:ED_lmg}

We compute $\chi_\mathrm{a}$ and the participation ratios by diagonalising the random LMG Hamiltonian of
Eq.~\eqref{eq:LMG_hamiltonian} for system sizes $ N = 303$ and $610$.
These system sizes correspond to the mean atom numbers realised in the experiment, which were determined from the dispersive shift $JN=g^2 N /\Delta_{\mathrm{ca}}$ measured at zero disorder ($W=0$), for each iteration of the measurement sequence [see end of Methods Sec.~\ref{s:atomprep}].
The effect of the atoms' thermal motion on the value of $g$ was taken into account for the conversion of the dispersive shifts into atom numbers, as well as for the matrix elements of the Hamiltonian.
Taking the mean atom number across all experimental runs, yields the system sizes quoted above.

We choose the random energy shifts $\epsilon_{i}$ in two different ways:
(i) from the incommensurate light-shift potential generating correlated quasi-random disorder
as discussed in the main text, and (ii) independent and identically distributed (i.i.d.) $\rho_\mathrm{a}$. For both cases, we find quantitative agreement of $\chi_\mathrm{a}$, and similarly of the PR, within numerical accuracy.

The Hamiltonian matrix is constructed with respect to the basis states $\ket{i} = \sigmah_i^+\ket{G}$ of the SEM, and diagonalised exactly.
In the absence of disorder, i.e., $\epsilon_{i} = 0 \, \forall i$, the diagonalisation is analytically tractable, and the eigenenergies are $\mathcal{E}_1 = -NJ/2$
and $\mathcal{E}_m = NJ/2$ for $m = 2, ..., N$.
From this follows that the zero-disorder ferromagnetic gap $\Delta_{\mathrm{FM}} \equiv \mathcal{E}_2 - \mathcal{E}_1 = JN$, as mentioned in the main text.
However, the presence of disorder mixes the Hamiltonian's zero-disorder eigenstates, necessitating the analysis through numerical diagonalisation.
Using the numerically determined eigenenergies and eigenstates, we compute the atomic susceptibility and PR from Eqs.~\eqref{e:atomic_susc}~and~\eqref{e:PRm}, respectively. 
We average these quantities with respect to $2000$ disorder realisations of the Hamiltonian, the results of which are illustrated in Figs.~\ref{fig:RLMG}~and~\ref{fig:PRbound} of the main text.
The corresponding variances are strongly suppressed, falling within the linewidths of the simulated data.

\pagebreak
\onecolumngrid
\newpage

\section{Supplementary Information}

\setcounter{page}{1} 
\setcounter{equation}{0} 
\setcounter{figure}{0} 
\renewcommand{\thepage}{S\arabic{page}} 
\renewcommand{\theequation}{S\arabic{equation}} 
\renewcommand{\thefigure}{S\arabic{figure}} 
\renewcommand{\figurename}{SUPPLEMENTARY FIG.} 

\subsection{Structure of the eigenstates in the near-resonant regime: multifractality and quasi-random potential}
\label{s:multifrac}

In this section, we discuss the structure of the eigenstates of the TC Hamiltonian \cref{eq:Hamiltonian} in the near-resonant regime ($\Delta_{\rm ca}=0$).
As recently found in Ref.~\cite{dubailLargeRandomArrowhead2022}, the eigenstates of the TC Hamiltonian with random uniformly distributed atomic energies are always multifractal, for any non-vanishing disorder strength. Here, we show that the same result is obtained with a correlated energy potential,
\begin{equation}
\label{e:AAHpotential}
\epsilon_{i} = \frac{W}{2}\cos (2\pi Q i),
\end{equation}
which is incommensurate to the lattice spacing [here we choose $Q=(\sqrt{5}-1)/2$]. This energy potential, in the presence of nearest-neighbour hopping and in the absence of a cavity mode, constitutes the well-studied Aubry--André--Harper (AAH) model, which has a localisation--delocalisation transition in the thermodynamic limit \cite{harperSingleBandMotion1955, aubryAnalyticityBreaking1980}. This is in contrast to the one-dimensional Anderson model, characterised by a random uniformly distributed energy potential with nearest-neighbour hopping, which has no transition and is always localised in the $N\to\infty$ limit \cite{andersonAbscenceDiffusionRandomLattices1958}. Moreover, the correlated potential in \cref{e:AAHpotential} with power-law hopping has a rich phase diagram, including localised, delocalised and multifractal phases, with mobility edges \cite{dengOneDimensionalQuasicrystals2019}.
Therefore, it is important to check whether this energy potential affects the results of Ref.~\cite{dubailLargeRandomArrowhead2022} regarding the multifractality of the TC Hamiltonian eigenstates.

\begin{figure*}[bth]
    \centering
    \includegraphics{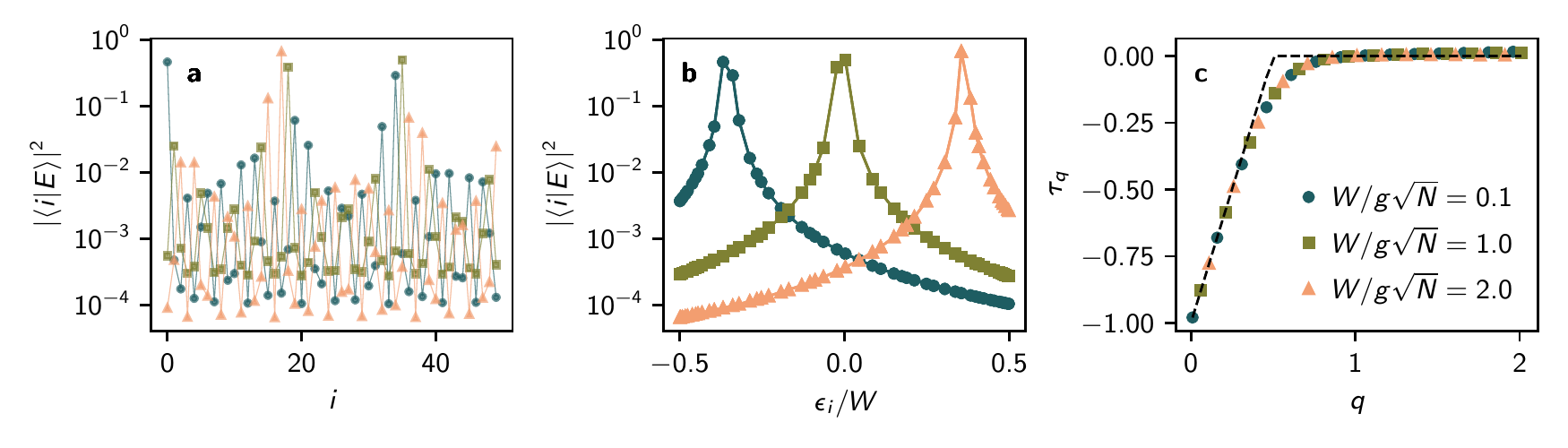}
    \caption{{\bf Multifractality of the eigenstates of the TC Hamiltonian near resonance.} {\bf a}, Squared amplitudes of a representative grey-state eigenfunction of the TC Hamiltonian \cref{eq:Hamiltonian} for $\Delta_{\rm ca}=0$ on the atoms $\ket{i}=\sigmah^+_i\ket{G}$, as a function of the atom index $i$.
    Here, $N=50$ and three different disorder strengths have been considered (see legend in panel {\bf c}). {\bf b}, Squared amplitudes of the same eigenfunctions shown in panel {\bf a}, here as a function of the atomic energies $\epsilon_{i}$ normalised by the disorder strength. {\bf c}, Multifractality of the eigenfunctions. The $\tau_q$ exponent has been determined from a power-law fit of  the generalised inverse participation ratio \cref{e:genIPR}, averaged over all the grey states, as a function of $N$.
        For the fitting procedure, several values of $N$ from $50$ to $2000$ have been considered.
        The dashed line is the analytical result obtained for the TC Hamiltonian with random uniformly distributed atomic energies in the thermodynamic limit~\cite{dubailLargeRandomArrowhead2022}. In all panels, we neglect the polaritons, and we normalised the grey-state eigenfunctions to their total probability on the atoms.}
    \label{fig:multi}
\end{figure*}

In \cref{fig:multi}{\bf a}, the squared amplitudes of some representative eigenfunctions $\ket{E}$ on the atoms $\ket{i}=\sigmah^+_i\ket{G}$ are shown as a function of the atom index $i$.
Different disorder strengths have been considered (see legend in panel {\bf c}), which are comparable to the range covered by the experiments (see \cref{fig:RTC} in the main text). As one can see, the eigenfunctions are characterised by few highly occupied atoms, and many atoms
with a small occupation probability, equally distributed in space. On the other hand, the occupation probabilities as a function of the atomic
energies have a power-law dependence, as shown in \cref{fig:multi}{\bf b}. These same features characterise the multifractal eigenstates of the TC Hamiltonian with random, uniformly distributed atomic energies, as shown in Ref.~\cite{dubailLargeRandomArrowhead2022}. 
Then, following Ref.~\cite{dubailLargeRandomArrowhead2022}, we quantify the multifractal behaviour of the eigenfunctions by analyzing the scaling of their generalised inverse participation ratio with the system size,
\begin{equation}
\label{e:genIPR}
{\rm IPR}_q(\ket{\psi}) = \sum_{i=1}^N |\langle i|\psi \rangle|^{2q} \propto N^{-\tau_q}~.
\end{equation}
The definition in \cref{e:genIPR} is valid for any normalised wavefunction $\ket{\psi}$, and here we consider the eigenfunctions $\ket{\psi} = \ket{E}$. The $\tau_q$ exponent is related to the multifractal dimension $D_q$ by the relation~\cite{eversAndersonTransitions2008}
\begin{equation}
\label{e:Dq}
D_q = d \frac{\tau_q}{q-1}~,
\end{equation}
where $d$ is the physical dimension of the system.
In \cref{fig:multi}{\bf c} the dependence of $\tau_q$ on the power $q$ is shown for different disorder strengths. The numerical results match very well the analytical results of Ref.~\cite{dubailLargeRandomArrowhead2022} (dashed line) up to some deviations around $q\approx 1/2$, due to the finite-size $N$. These results confirm that the eigenfunctions are multifractal with the correlated potential in \cref{e:AAHpotential} that has been realised in the experiment described in the main text.

\subsection{Modelling via Lindblad equation}
\label{s:lindblad}

Here, we provide further details on the model used to derive the relation between the atomic susceptibility $\chi_\mathrm{a}$ and the population of the auxiliary state $P_\mathrm{A}(t)$, given in Eq.~\eqref{e:me_popA} of the Methods.

We model the experimental sequence of Secs.~\ref{s:atomprep}--\ref{s:measure_atom_excitations} via a Lindblad master equation, which allows us to derive an equation of motion for $P_\mathrm{A}(t)$ in terms of $\dsa$.
We work in a rotating frame generated by $N \Delta_{\mathrm{pa}} \hat{S}^z$, such that the Hamiltonian of Eq.~\eqref{e:Heff} is time-independent, i.e., $\hat{\mathcal{V}}(t) \to \hat{\mathcal{V}} =  \hat{\mathcal{V}}(0)$.
The Lindblad equation is then given by
\begin{equation}\label{e:me}
\partial_t \hat{\rho}(t) = -i\comm{ \hat{H}_\mathrm{LMG} - N \Delta_{\mathrm{pa}} \hat{S}^z + \hat{\mathcal{V}} }{ \hat{\rho}(t) }
+ \! \bigl(  \mathcal{D}[\Gamma_\mathrm{g}; \{ \sigmah_i^- \}] \! + \!  \mathcal{D}[\Gamma_\mathrm{a}; \{ \ketbra{a}{e}_i \}] \bigr) \hat{\rho} (t) ,
\end{equation}
where the superoperators $\mathcal{D}[\gamma'; \lbrace \hat{L}_i \rbrace] \hat{\rho}(t) \equiv \gamma' \sum_{i=1}^N \left( \hat{L}_i \hat{\rho}(t) \hat{L}_i^\dagger - \acomm{\hat{L}_i^\dagger \hat{L}_i}{\hat{\rho}(t)} / 2 \right)$ describe dissipation at a rate $\gamma'$, due to jump processes generated by $\lbrace \hat{L}_i \rbrace$.
Specifically, the superoperators with rates $\Gamma_\mathrm{g}$ and $\Gamma_\mathrm{a}$ describe spontaneous decay of atoms from $\ket{e}$ to $\ket{g}$ and $\ket{a}$, respectively.
The decay rates $\Gamma_\mathrm{g},\Gamma_\mathrm{a}$ are branching ratios of the natural linewidth $\Gamma=5.8\times 2\pi \, \mathrm{MHz}$ of the D$2$ line of $^6\mathrm{Li}$, i.e., $\Gamma_\mathrm{g} + \Gamma_\mathrm{a} = \Gamma$.
Spontaneous emission from $\ket{a}$ to $\ket{g}$ can be neglected on the timescales of the experiment, as for $^6\mathrm{Li}$ it is forbidden by selection rules.

The presence of a single atom in state $\ket{a}$ drastically suppresses the transmission signal, due to the cavity's high cooperativity.
The experiment must therefore be executed in a regime where at most one atom is in state $\ket{a}$ so as to avoid saturation of the transmission signal (see \cref{fig:extended_seq}\textbf{c}).
We therefore project the dynamics of Eq.~\eqref{e:me} onto the Hilbert subspace with at most one excitation and at most one atom in the auxiliary state $\ket{a}$.
We then utilise a separation of scales to derive the equation of motion for $P_\mathrm{A}(t) \equiv \sum_{i=1}^N \mel{a_i}{\hat{\rho}(t)}{a_i}$ (where $\ket{a_i} \equiv \dyad{a}{g}_i \ket{G}$):
Within the time domain ${t \gg (\Gamma/2)^{-1}}$, all coherences as well as the SEM populations can be adiabatically eliminated from the rate equations of the remaining populations  $p_G(t)$ and $\lbrace \mel{a_i}{\hat{\rho}(t)}{a_i} \rbrace_{i=1}^N$.
Doing so, one finds that $\partial_t p_G(t) = - \partial_t P_\mathrm{A}(t)$ (conservation of atomic population), and to lowest order in $(\Gamma/2)^{-1}$
\begin{equation}\label{e:rate_eqs}
\partial_t p_G(t) \,=\, - \Gamma_\mathrm{a} \sum_{m \in \mathrm{SEM} } \frac{\abs{\mathcal{V}_{mG}}^2}{(\Gamma / 2)^2 + (E_{mG} - \Delta_{\mathrm{pa}})^2} p_G(t)  
\, = \,
- \frac{ \Gamma_\mathrm{a}}{(\Gamma/2)^2}  \left\lvert \frac{g \Omega_{\mathrm{p}}}{\Delta_{\mathrm{ca}}} \right\rvert^2 \chi_{\mathrm{a}}(\Delta_{\mathrm{pa}})  p_G(t) .
\end{equation}
To obtain the relation to $\dsa$, as defined by Eq.~\eqref{e:atomic_susc}, we have used that (within the rotating frame of $N \Delta_{\mathrm{pa}} \hat{S}^z$) the matrix elements $\mathcal{V}_{mG}$ follow from Eq.~\eqref{e:Heff_V} as ${ \mathcal{V}_{mG} = \frac{g \sqrt{N} \Omega_{\mathrm{p}}}{\Delta_{\mathrm{ca}}}  \mel{m}{\hat{S}^+}{G} }$, and we have identified the Lorentzian response of Eq.~\eqref{e:lorenzian_response}, with linewidth ${\gamma = \Gamma/2}$.

For the initial conditions $p_G(0)=1, P_\mathrm{A}(0)=0$, we finally obtain the relation stated in Eq.~\eqref{e:me_popA}, 
\begin{equation}
P_\mathrm{A}(t_\mathrm{meas}) =  1 - \exp( - \frac{ \Gamma_\mathrm{a}}{(\Gamma/2)^2}  \left\lvert \frac{g \Omega_{\mathrm{p}}}{\Delta_{\mathrm{ca}}} \right\rvert^2 \chi_{\mathrm{a}}(\Delta_{\mathrm{pa}})   t_\mathrm{meas})  .
\end{equation}
For further discussions of the properties of this relation, we refer the reader to Sec.~\ref{s:meas_atomic_sus} of the Methods.

\subsection{Finite-size scaling of the minimal ferromagnetic gap in the large-detuning regime}
\label{s:SM_gapclosing}
\begin{figure}[htb]
    \includegraphics{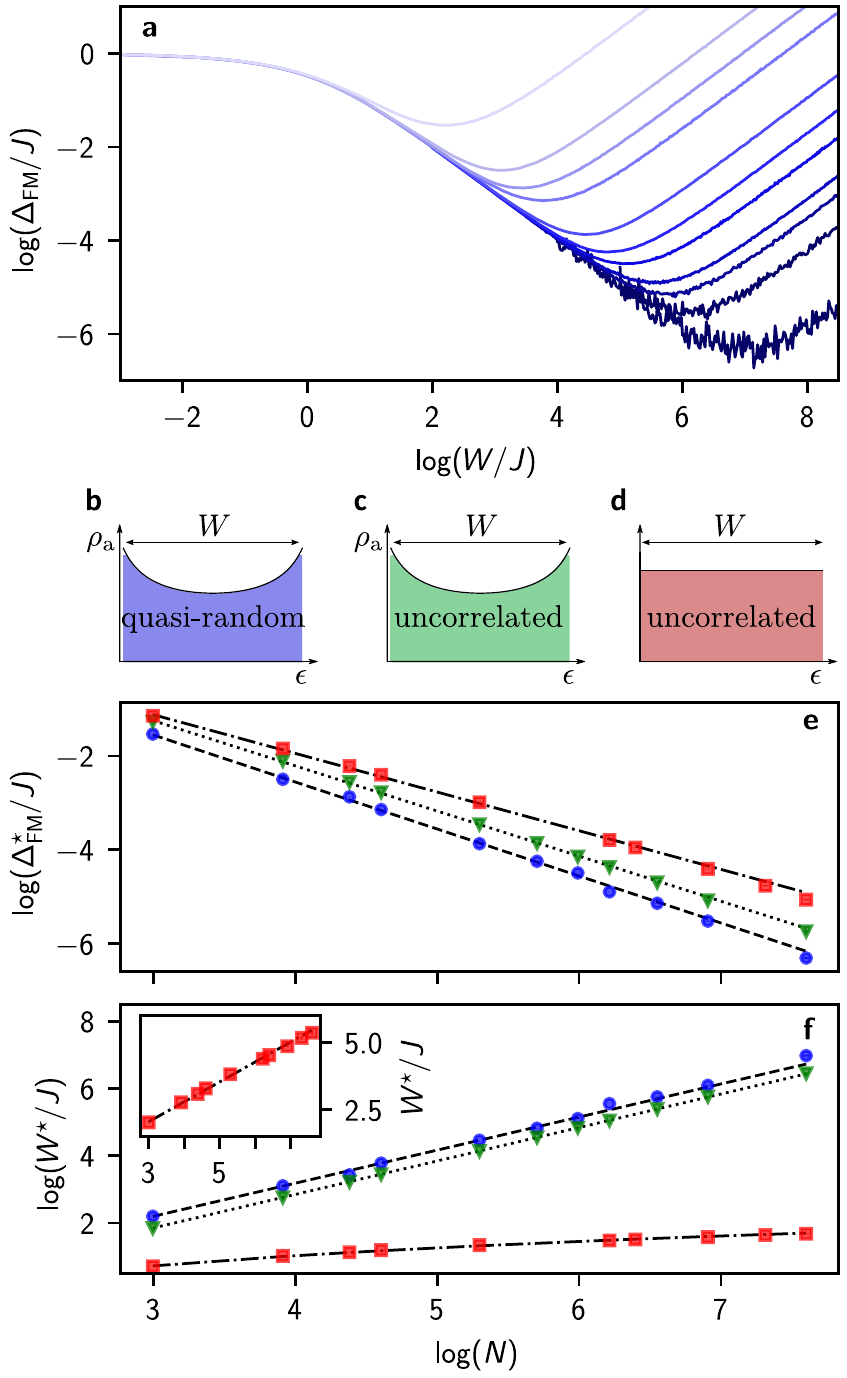}
    \caption{\label{fig:ext_gapclosing}
        \textbf{Finite-size scaling of the minimal ferromagnetic gap of the random LMG model.}
        \textbf{a}, Disorder averaged ferromagnetic gap $\Delta_{\mathrm FM} /J$ of the
        Kac-normalised random LMG Hamiltonian, as a function of disorder strength $W /J$.
        Only data for the quasi-disordered case is shown as a representative.
        Lighter to darker shades of blue correspond to increasing $N$ from $20$ to $2000$.
        For each $N$, the minimal gap $\Delta_{\mathrm FM}^{\star} /J$ and its location $W^{\star} /J$ are determined
        from a parabolic fit to the corresponding curve.
        \textbf{b}--\textbf{d}, Schematic distributions of the considered disorders:
        quasi-random and i.i.d.\
        $\epsilon_i$ from $\rho_{\mathrm{a}}$, and i.i.d.\ $\epsilon_i$ from uniform distribution, left to right.
        \textbf{e}~and~\textbf{f}, $\log(\Delta_{\mathrm FM}^{\star} / J)$ and $\log(W^{\star} / J)$ versus $\log(N)$, respectively, for the quasi-random $\rho_{\mathrm{a}}$
        (blue circles), uncorrelated $\rho_{\mathrm{a}}$ (green triangles),
        and uniform (red squares) disorder distributions.
        The errors in the parabolic fits are smaller than the size of the markers.
        From linear fits to the data points for the quasi-random (dashed) and the uncorrelated (dotted) $\rho_{\mathrm{a}}$, we find $W^{\star} /J \propto N^{\alpha}$ with $\alpha \approx 0.98$ and $0.99$, respectively, indicating linear dependence. 
        For the uniform disorder distribution, we fit a logarithmic curve (dot-dashed), which suggests that $W^{\star} /J \propto \log(N)$.
        This is verified by the linear fit (dot-dashed) in the inset of \textbf{e}, where the $y$-axis depicts $W^\star / J$ instead of $\log(W^\star /J)$.
        In contrast, the minimal ferromagnetic gap $\Delta_{\mathrm FM}^{\star} /J$ scales as $N^{-\beta}$ with
        $\beta \approx 1.00$, $0.96$, and $0.82$ for the considered disorders, respectively.
    }
\end{figure}

Here, we analyse the finite-size scaling of the minimal ferromagnetic gap $\Delta_{\mathrm{FM}}$ of the disordered LMG model in Eq.~\eqref{eq:LMG_hamiltonian}.
For this, we first Kac normalise the all-to-all spin-exchange interaction term in the Hamiltonian by $N^{-1}$, which renders the model extensive.
In contrast to the experimental scenario, Kac normalisation is necessary to theoretically analyse any critical behaviour stemming from the competition between different terms in the Hamiltonian.
Under this rescaling, the zero-disorder gap is $\Delta_{\mathrm{FM}}/J =  1$, which decreases to a minimal value $\Delta_{\mathrm{FM}}^{\star}/J$ as the disorder strength $W/J$ is increased.
We denote the disorder strength at which this minimum is realised as $W^{\star}/J$. 
The minimal gap is indicative of significant changes in the ground-state properties in a finite-size system.
We perform a finite-size scaling of $\Delta_{\mathrm{FM}}^{\star}/J$ and $W^{\star}/J$ for the system with (i) quasi-random disorder [sampled from the correlated energy potential of Eq.~\eqref{e:AAHpotential}], and compare it with (ii) uncorrelated disorder with distribution $\rho_{\mathrm{a}}$ (as defined in Sec.~\ref{s:model} of the main text), and (iii) uniform
distributions (see Fig.~\ref{fig:ext_gapclosing}).

The dependence of $\Delta_{\mathrm{FM}}/J$ on $W/J$ is shown in Fig.~\ref{fig:ext_gapclosing}\textbf{a}
for the correlated disorder, which (adapting for different scalings, see below) is representative also of the two other studied cases.
The minimal gap $\Delta_{\mathrm FM}^{\star}/J$,
and its location $W^{\star}/J$, are determined by fitting a parabola to the minimum of
the curves.
For all three disorder distributions, the gap $\Delta_{\mathrm{FM}}^{\star}/J$ decreases as $N^{-\beta}$ 
(see Fig.~\ref{fig:ext_gapclosing}\textbf{e}), indicating that the gap disappears in the thermodynamic
limit. 
The gap location $W^{\star}/J$, however, scales linearly with system-size
($\propto N^{\alpha}$ with $\alpha\approx 1$) for the uncorrelated and correlated
$\rho_{\mathrm{a}}$ (see Fig.~\ref{fig:ext_gapclosing}\textbf{f}).
In contrast, $W^{\star}/J\propto \log N$ for the uncorrelated uniform
disorder (inset of Fig.~\ref{fig:ext_gapclosing}\textbf{f}), which is consistent with the vanishing Richardson's superconducting gap, which was
estimated for uniform disorder from the mean level
spacing \cite{celardoShieldingLocalization2016}.
These findings are in agreement with the generic behaviour of the critical disorder strength $W_\mathrm{c}$ for the Anderson localisation transition in
models with a  high connectivity, which increases with the number of connections \cite{Abou_Chacra_1973, mirlinDistributionLocalDensities1994, tarquiniCriticalPropertiesAndersonLocalization2017,royLocalizationCertainGraphs2020}.
For example, in a $d$-dimensional hypercube with coordination number $z\sim 2^d$,
one finds $W_c\propto d \log d$ \cite{tarquiniCriticalPropertiesAndersonLocalization2017}.
By visualising the all-to-all
connectivity realised in the large-detuning regime as a hypercube with dimension $d\sim N$, we can expect $W_\mathrm{c}\propto \log N$
to leading order. This similarity indicates the significant change in
localisation properties that a finite system experiences around $W^{\star}/J$, which is also supported by the
decreasing trend in the PR (similar to Fig.~\ref{fig:PRbound}). Correlations in the disorder tend to delocalise the system more, consistent with our finding in Fig.~\ref{fig:ext_gapclosing}\textbf{f}. 
In the thermodynamic limit, however, the infinitely-connected system does not support a localisation transition.

\end{document}